\pdfoutput=1
\documentclass[11pt,twoside,a4paper,cmspaper,final,collab]{cms-tdr}
\def\svnVersion{518151b}\def\svnDate{2024/11/06}\def\cmsCernNoTag{CERN-EP-2024-167}\def\cmsCernDate{\today}\def\cmsMessage{Published in Physical Review D as \href{http://dx.doi.org/10.1103/PhysRevD.110.092002}{\doi{10.1103/PhysRevD.110.092002}.}}
\begin{document}\cmsNoteHeader{HIN-23-013}

\newcommand{\TJPsi}{\JPsi\JPsi+\PX}
\newcommand{\SPS} {\ensuremath{(\JPsi\JPsi\PGf)_\text{SPS}}\xspace}
\newcommand{\DPS} {\ensuremath{(\JPsi\JPsi)_\text{SPS}+\PGf}\xspace}
\newcommand{\DPSt}{\ensuremath{(\JPsi\PGf)_\text{SPS}+\JPsi}\xspace}
\newcommand{\TPS}{\ensuremath{\JPsi_\text{SPS}+\JPsi_\text{SPS}+\PGf_\text{SPS}}\xspace}
\newcommand{\pp}{\ensuremath{\Pp\Pp}\xspace}
\newcommand{\pPb}{\ensuremath{\Pp\mathrm{Pb}}\xspace}
\newcommand{\mumu}{\ensuremath{\PGmp\PGmm}\xspace}
\newcommand{\jpsimumu}{\ensuremath{\JPsi\to\mumu}\xspace}
\newcommand{\jpsiee}{\ensuremath{\JPsi\to\Pep\Pem}\xspace}
\newcommand{\KK}{\ensuremath{\PKp\PKm}\xspace}
\newcommand{\massmumu}{\ensuremath{m_{\mumu}}\xspace}
\newcommand{\massKK}{\ensuremath{m_{\KK}}\xspace}
\newcommand{\psio}{\ensuremath{\psi_1}\xspace}
\newcommand{\psid}{\ensuremath{\psi_2}\xspace}
\newcommand{\xsec}{\sigma(\pPb\to\TJPsi)}
\newcommand{\yield}{\ensuremath{8.5 \pm 3.4}\xspace}
\newcommand{\ntjpsisig}{\ensuremath{N_\text{sig}}\xspace}
\newcommand{\EPOS} {\textsc{epos-lhc}\xspace}
\newcommand{\brjmm}{\ensuremath{\mathcal{B}_{\JPsi\to\mumu}}\xspace}
\newcommand{\xsecbrfull}{\ensuremath{\xsec = \frac{\ntjpsisig}{\epsilon\,\Lumi_\text{int}\,\mathcal{B}^{2}_{\JPsi\to\mumu}}}\xspace}
\newcommand{\sigmaeff}{\ensuremath{\sigma_\text{eff}}\xspace}
\newcommand{\sigmaeffdps}{\ensuremath{\sigma_\text{eff}}\xspace}
\newcommand{\sigmaeffdpspPb}{\ensuremath{\sigma_\text{eff,\pPb}}\xspace}
\newcommand{\sigmaeffdpspPbresult}{\ensuremath{\sigma_\text{eff,\pPb}=0.53^{+\infty}_{-0.2}\unit{b}}\xspace}
\newcommand{\expm}{\ensuremath{\,\text{(exp)}}\xspace}
\newcommand{\pA}{\ensuremath{\Pp\mathrm{A}}\xspace}
\newcommand{\DeltaPhi}{\ensuremath{\Delta\phi}\xspace}
\newcommand{\DeltaY}{\ensuremath{\Delta y}\xspace}
\newcommand{\jpsio}{\ensuremath{\JPsi_{1}}\xspace}
\newcommand{\jpsit}{\ensuremath{\JPsi_{2}}\xspace}

\newcommand{\helaconia}{\textsc{HELAC-Onia}\xspace}

\newcommand{\result}{22.0\pm 8.9 \stat \pm 1.5 \syst \unit{nb}}
\newcommand{\resultDPS}{5.4\pm 6.2\stat \pm 0.4 \syst \unit{nb}}
\newcommand{\resultSPS}{16.5\pm 10.8\stat \pm 0.1 \syst \unit{nb}}

\hyphenation{resolution}

\cmsNoteHeader{HIN-23-013}

\ifthenelse{\boolean{cms@external}}{\providecommand{\cmsLeft}{top\xspace}}{\providecommand{\cmsLeft}{left\xspace}}
\ifthenelse{\boolean{cms@external}}{\providecommand{\cmsRight}{bottom\xspace}}{\providecommand{\cmsRight}{right\xspace}}
\newlength{\cmsFigSize}
\ifthenelse{\boolean{cms@external}}{\setlength{\cmsFigSize}{0.30\textwidth}}{\setlength{\cmsFigSize}{0.45\textwidth}}

\title{Observation of double \texorpdfstring{\JPsi}{J/psi} meson production in \texorpdfstring{\pPb}{pPb} collisions at \texorpdfstring{\sqrtsNN = 8.16\TeV}{sqrt(s[NN]) = 8.16 TeV}}

\date{\today}

\abstract{
The first observation of the concurrent production of two \JPsi mesons in proton-nucleus collisions is presented. The analysis is based on a proton-lead (\pPb) data sample recorded at a nucleon-nucleon center-of-mass energy of 8.16\TeV by the CMS experiment at the CERN LHC and corresponding to an integrated luminosity of 174.6\nbinv. The two \JPsi mesons are reconstructed in their \mumu decay channels with transverse momenta $\pt>6.5\GeV$ and rapidity $\abs{y}<2.4$. Events where one of the \JPsi mesons is reconstructed in the dielectron channel are also considered in the search. The $\pPb\to\TJPsi$ process is observed with a significance of 5.3 standard deviations. The measured inclusive fiducial cross section, using the four-muon channel alone, is ${\sigma(\pPb\to\TJPsi)= \result}$. A fit of the data to the expected rapidity separation for pairs of \JPsi mesons produced in single (SPS) and double (DPS) parton scatterings yields ${\sigma^{\pPb\to\TJPsi}_\text{SPS}=\resultSPS}$ and ${\sigma^{\pPb\to\TJPsi}_\text{DPS}=\resultDPS}$, respectively. This latter result can be transformed into a lower bound on the effective DPS cross section, closely related to the squared average interparton transverse separation in the collision, of $\sigmaeff>1.0\unit{mb}$ at 95\% confidence level.
}

\hypersetup{
pdfauthor={CMS Collaboration},
pdftitle={Observation of double J/psi meson production in pPb collisions at sqrt(s[NN]) = sqrt(s[NN]) = 8.16 TeV},
pdfsubject={CMS},
pdfkeywords={QCD, double parton scatterings, vector mesons}}

\maketitle 

\section{Introduction}

High-energy collisions of protons and nuclei at the CERN LHC are characterized by multiple interactions of their underlying partonic constituents (quarks and gluons). Such multiple parton interactions can lead to the simultaneous production of several particles with values of transverse momentum and/or mass above a few \GeV, much larger than the typical nonperturbative $\mathcal{O}(0.2\GeV)$ scale of quantum chromodynamics (QCD). Research on multiparton interactions has attracted an increasing interest at hadron colliders as a means to probe the generalized parton densities of the proton, to determine the unknown energy evolution of the partonic transverse profile of the proton, and to study the role of partonic correlations (in position and momentum phase space, as well as in flavor, color, and spin quantum numbers) in the hadronic wave functions~\cite{Diehl:2017wew,Bartalini:2018qje}. A good understanding of multiple hard scatterings of partons is also crucial to describe the continuum backgrounds in studies of rare standard model resonance decays, as well as in searches for new physics, in final states where various heavy particles are produced~\cite{CMS:2019wch,Sirunyan:2020txn,CMS:2023owd}.

Following a purely geometric approach, \ie, ignoring any parton correlations, the probability to produce $n$ high-\pt particles in a given hadron-hadron collision is proportional to the $n^\text{th}$-product of the probabilities to independently produce each of them~\cite{dEnterria:2017yhd}. Thus, the probability to produce two high-\pt particles in a double parton scattering (DPS) would scale with the square of the corresponding single parton scattering (SPS) probabilities. The occurrence of DPS processes is therefore more likely for final states with large SPS cross sections, as is the case for quarkonium mesons compared to rarer heavy particles such as electroweak bosons~\cite{Chapon:2020heu}. Under this approximation, the cross section to produce two charmonium mesons, \psio and \psid, via DPS in a proton-proton (\pp) collision can be expressed as the product of the SPS cross sections for the production of each individual meson (calculated using perturbative QCD or directly measured using collision data) normalized by an effective cross section ($\sigmaeff$) to ensure the proper units of the ratio,
\begin{equation}
\sigma^{\pp \to \psio\,\psid+\PX}_\text{DPS} = \left(\frac{\mathfrak{m}}{2}\right)\, \frac{\sigma^{\pp \to \psio+\PX}_\text{SPS} \, \sigma^{\pp\to \psid+\PX}_\text{SPS}}{\sigmaeffdps}.
\label{eq:pocketDPS}
\end{equation}
Here, $\mathfrak{m}$ is a combinatorial factor to avoid double counting the same process, $\mathfrak{m}=1\,(2)$ if $\psio=\psid$ ($\psio\neq\psid$). In a purely geometric approach, the $\sigmaeffdps$ value can be determined via 
\begin{equation} 
\sigmaeff=\left[ \int \rd^2\mathbf{b}\,T^2(\mathbf{b})\right]^{-1},
\label{eq:sigmaeffDPS}
\end{equation}
where $T(\mathbf{b})$ is the \pp transverse overlap function at the vector impact parameter $\mathbf{b}$ of the collision~\cite{dEnterria:2017yhd}. In terms of the transverse parton density of the proton, $\rho(\mathbf{b})$, the overlap function is given by $T(\mathbf{b})=\int \rho(\mathbf{b_1}) \rho(\mathbf{b_1 -b})\rd^2\mathbf{b_1}$, and the $\sigmaeff$ parameter can be interpreted as the squared average interparton distance in the \pp collision where the two hard scatterings take place. Thus, a smaller value of $\sigmaeff$ implies larger DPS yields. For the typical proton transverse density profiles implemented in common Monte Carlo (MC) event generators used at collider energies, \eg, \PYTHIA~\cite{Sjostrand:2017cdm} or \HERWIGpp~\cite{Seymour:2013qka}, $\sigmaeffdps\approx20\mbox{--}30\unit{mb}$. These values are about a factor of two larger than those experimentally derived via Eq.~(\ref{eq:pocketDPS}) for a variety of different pairs of particles measured at the Tevatron and the LHC~\cite{Bartalini:2018qje}. This discrepancy points to the presence of various types of parton correlations in the proton~\cite{Blok:2017alw} not accounted for in the purely geometrical approach, as represented in Eq.~(\ref{eq:sigmaeffDPS}). In addition, the experimentally extracted values of $\sigmaeff$ seemingly depend on the particular final state under consideration, with multi-quarkonium production processes yielding lower values, $\sigmaeffdps\approx 5\mbox{--}10\unit{mb}$ from $\JPsi\JPsi$~\cite{Aaij:2011yc,Abazov:2014qba,Khachatryan:2014iia,Aaboud:2016fzt,Aaij:2016bqq}, $\JPsi\JPsi\JPsi$~\cite{CMS:2021qsn}, $\JPsi\Upsilon$~\cite{Abazov:2015fbl}, and $\Upsilon\Upsilon$~\cite{Khachatryan:2016ydm,Sirunyan:2020txn} studies, than the $\sigmaeffdps\approx 15\unit{mb}$ obtained from final states involving pairs of high-\pt jets and/or electroweak bosons~\cite{ATLAS:2013aph,CMS:2013huw,ATLAS:2016rnd,CMS:2017han,Aaboud:2018tiq,CMS:2019jcb,CMS:2021lxi,CMS:2022pio}. Such process-dependent $\sigmaeffdps$ determinations appear consistent with flavor-dependent correlations that follow the different quark or gluon densities in the parton distribution functions (PDFs) probed at varying values of the momentum fraction by each scattering~\cite{Huayra:2023gio}.

\begin{figure*}[htpb!]
\centering
    \includegraphics[width=.95\textwidth]{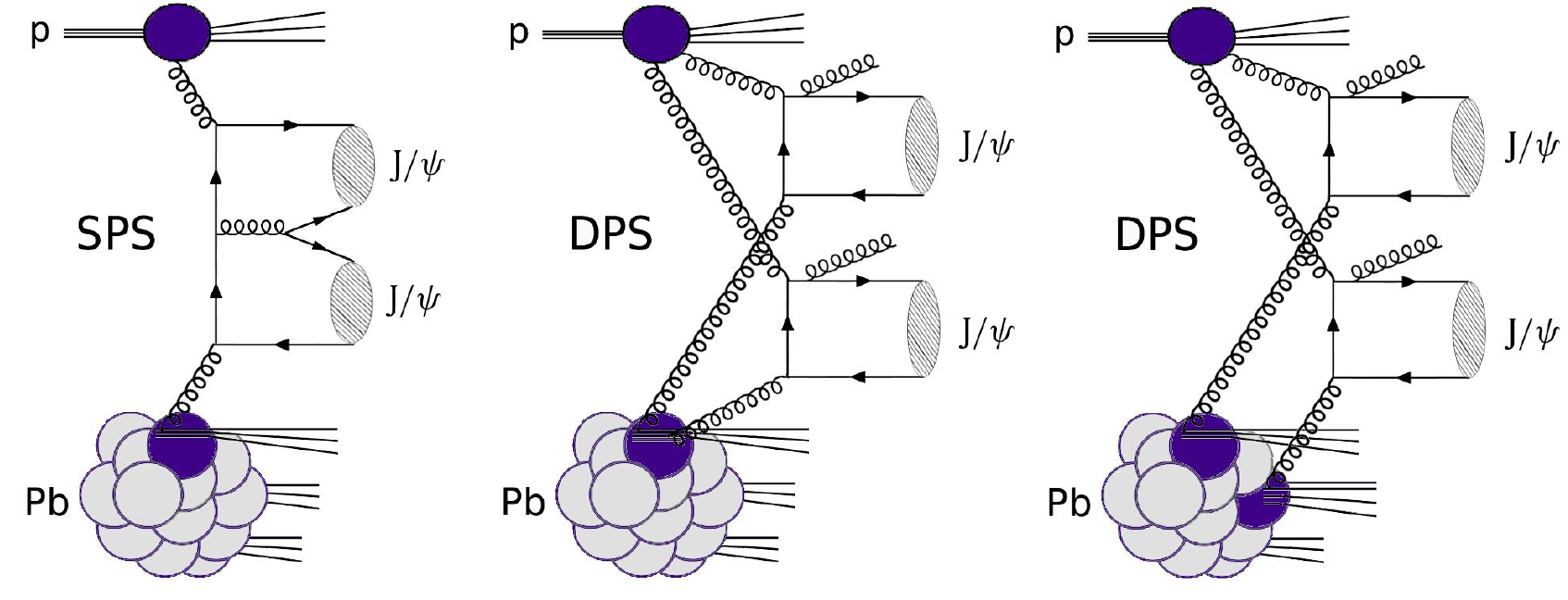}
    \caption{Representative diagrams of SPS (left) and DPS (center and right) contributions to the production of two \JPsi mesons in $\pPb$ collisions. \label{fig:diags}}
\end{figure*}

To advance understanding of multiple hard scatterings, it has been suggested to study DPS processes in proton-nucleus (\pA) collisions~\cite{Strikman:2001gz,Cattaruzza:2004qb,dEnterria:2012jam} in which the large transverse parton density of the nucleus significantly enhances the DPS contributions compared to the \pp case. Measurements of DPS processes in \pA collisions have also been proposed as a means to shed light on the impact-parameter dependence of nuclear PDFs (nPDFs)~\cite{Shao:2020acd}. For \pPb collisions, the equivalent of Eq.~(\ref{eq:pocketDPS}) for the inclusive production of two \JPsi mesons is
\begin{equation}
\sigma^{\pPb \to \TJPsi}_\text{DPS} = \left(\frac{1}{2}\right)\, \frac{\sigma^{\pPb \to \JPsi+\PX}_\text{SPS} \, \sigma^{\pPb\to \JPsi+\PX}_\text{SPS}}{\sigmaeffdpspPb},
\label{eq:pocketDPSpPb}
\end{equation}
where the effective \pPb\ cross section in the denominator accounts now for two possible DPS contributions coming from collisions of partons of the ``projectile'' proton with partons of one or two nucleons of the ``target'' lead ion (center and right diagrams of Fig.~\ref{fig:diags}, respectively). Under the same assumptions of factorization of hard-scattering probabilities that lead to Eq.~(\ref{eq:sigmaeffDPS}) in the \pp case, the value of $\sigmaeffdpspPb$ can be connected to that of $\sigmaeffdps$ through the relationship~\cite{dEnterria:2012jam,dEnterria:2017yhd}
\begin{equation}
\sigmaeffdpspPb= \frac{A\,\sigmaeffdps}{1+\sigmaeffdps\,F_{\pPb}/A}\,,
\label{eq:sigmaeffdpspA} 
\end{equation}
where $A = 208$ is the Pb mass number, and the quantity 
${F_{\pPb}} = (A-1)/A\,\int \rd^2\mathbf{b}\,T^2(\mathbf{b})\approx A^{1/3}/14\pi\unit{mb}^{-1}$, similarly to Eq.~(\ref{eq:sigmaeffDPS}), is the integral over impact parameter of the squared \pPb\ transverse profile, which can be determined through a Glauber MC model~\cite{dEnterria:2020dwq}. In this approximation, the value of $\sigmaeffdpspPb$ is expected to be about half the total hadronic \pPb\ cross section that amounts to $\sigma_{\pPb}\approx 2\unit{b}$ at the LHC~\cite{CMS:2015nfb,Loizides:2017ack}. Using Eq.~(\ref{eq:sigmaeffdpspA}), measurements of the DPS cross sections in \pPb\ collisions can therefore provide alternative extractions of the $\sigmaeffdps$ parameter, independently of \pp data. Such studies with \pPb\ collisions extend the physics reach of the heavy ion program at the LHC~\cite{CMS:2024krd}.

A first study of DPS in \pPb collisions was carried out by the LHCb experiment through measurements of the simultaneous production of different pairs of charmed hadrons at nucleon-nucleon center-of-mass energy $\sqrtsNN =8.16\TeV$~\cite{LHCb:2020jse}. This paper reports the measurement of the concurrent production of two \JPsi vector mesons in $\pPb$ collisions at the same center-of-mass energy. This process has not been previously observed and is of interest to clarify the SPS production of multiple quarkonium states~\cite{Chapon:2020heu,Lansberg:2019adr}, to improve our understanding of DPS processes~\cite{dEnterria:2017yhd,Blok:2017alw}, as well as to study nPDF properties~\cite{Shao:2020acd}. The generic types of partonic scatterings contributing to prompt $\TJPsi$ production are schematically shown in Fig.~\ref{fig:diags}. In SPS processes (Fig.~\ref{fig:diags}, left), the two vector mesons are produced by a single pair of interacting partons, whereas two different pairs of partons produce the final $\TJPsi$ state in DPS (Fig.~\ref{fig:diags}, center and right). The relatively high mass ($m_{\JPsi} = 3.1\GeV$), combined with the experimental requirement of $\pt^{\JPsi}>6.5\GeV$, ensure that both charmonium mesons are produced in hard partonic scatterings with virtualities above $Q \approx \sqrt{\smash[b]{m_{\JPsi}^2+\pt^2}} \approx 7\GeV$, which can be well described theoretically with perturbative calculations based, \eg, on the nonrelativistic QCD approach~\cite{Bodwin:1994jh}.

This analysis studies DPS in double charmonium production in \pPb\ collisions at the LHC, following an approach similar to that used in the triple-\JPsi observation in \pp collisions~\cite{CMS:2021qsn}. Exploiting the $\pPb$ data sample collected at $\sqrtsNN = 8.16 \TeV$ in 2016, we measure the fiducial cross section for the rare final state of interest. The $\TJPsi$ cross section is measured in the fiducial volume of the detector, defined in Table~\ref{tab:fiducialregion}, correcting for experimental inefficiencies. There is no extrapolation of the cross section for the final state beyond the fiducial acceptance, thus avoiding potential systematic uncertainties coming, \eg, from the polarization of the \JPsi mesons. The tabulated results of this study are provided in a HEPData record~\cite{HEPData}.

\begin{table}[htpb!]
  \centering
  \topcaption{Definition of the fiducial phase space for the $\pPb\to\TJPsi$ cross section measurement in $\pPb$ collisions at $\sqrtsNN=8.16\TeV$.}
  \begin{scotch}{ll} 
  Particle & Fiducial requirement \\ \hline
  Muons     & $\pt>3.4\GeV$ for $\abs{\eta} < 0.3$  \\
	              & $\pt>3.3\GeV$ for $0.3 < \abs{\eta} < 1.1$ \\
            	  & $\pt>5.5\mbox{--}2.0 \abs{\eta}\GeV$ for $1.1 < \abs{\eta} < 2.1$ \\
            	  & $\pt>1.3\GeV$ for $2.1 < \abs{\eta} < 2.4$  \\ [1.0ex]
  \JPsi mesons & $\pt>6.5\GeV$ and $\abs{y}<2.4$  \\
  \end{scotch}
\label{tab:fiducialregion}
\end{table}

\section{The CMS detector and data selection}

The CMS apparatus~\cite{CMS:2008xjf} is a multipurpose, nearly hermetic detector, designed to trigger on~\cite{CMS:2020cmk,CMS:2016ngn} and identify electrons, muons, photons, and charged and neutral hadrons~\cite{CMS:2020uim,Sirunyan:2018fpa,CMS:2014pgm}. A global ``particle-flow'' algorithm~\cite{CMS:2017yfk} aims to reconstruct all individual particles in an event, combining information provided by the all-silicon inner tracker and by the crystal electromagnetic and brass-scintillator hadron calorimeters, operating inside a 3.8\unit{T} superconducting solenoid, with data from the gas-ionization muon detectors embedded in the flux-return yoke outside the solenoid. The silicon tracker measures charged particles within the pseudorapidity range $\abs{\eta} < 2.5$. During the LHC running period when the data used in this Letter were recorded, the silicon tracker consisted of 1440 silicon pixel and 15\,148 silicon strip detector modules. For nonisolated particles of $1 < \pt < 10\GeV$ and $\abs{\eta} < 1.4$, the track resolutions are typically 1.5\% in \pt and 25--90 (45--150)\mum in the transverse (longitudinal) impact parameter~\cite{CMS:2014pgm}. The electromagnetic calorimeter (ECAL) consists of 75\,848 lead tungstate crystals, which provide coverage in pseudorapidity $\abs{\eta} < 1.48$ in a barrel region and $1.48 < \abs{\eta} < 3.0$ in two endcap regions. Muons are measured in the pseudorapidity range $\abs{\eta} < 2.4$, with detection planes made using three technologies: drift tubes, cathode strip chambers, and resistive plate chambers. The forward hadron (HF) calorimeter extends the pseudorapidity coverage provided by the barrel and endcap detectors, and uses steel as an absorber and quartz fibers as the sensitive material. The two halves of the HF are located 11.2\unit{m} from the interaction region, one on each end, and together they provide coverage in the range $3.0 < \abs{\eta} < 5.2$. They also serve as luminosity monitors. A more detailed description of the CMS detector, together with a definition of the coordinate system used and the relevant kinematic variables, can be found in Ref.~\cite{CMS:2008xjf}.

Events are filtered using a two-tiered trigger system~\cite{CMS:2016ngn}. The first level, composed of custom hardware processors, uses information from the calorimeters and muon detectors~\cite{CMS:2020cmk}. The second level, known as the high-level trigger, consists of a farm of processors running a version of the full event reconstruction software optimized for fast processing, and reduces the event rate to around 1\unit{kHz} before data storage~\cite{CMS:2016ngn}.

The events are collected with a trigger requiring two muon candidates found in the muon detectors at level-1. In order to maximize the detection efficiency~\cite{CMS:2024qgs}, no explicit selection is made on their transverse momentum or pseudorapidity. During the 2016 \pPb run, this trigger was operated without any prescale, recording an integrated luminosity of 174.6\nbinv.

For the offline analysis, events have to pass a set of selection criteria designed to reject events from background processes (beam-gas collisions) as described in Ref.~\cite{Khachatryan:2016odn}. Events are required to have at least one reconstructed interaction vertex, formed by two or more associated tracks, with a distance from the center of the nominal interaction region of less than 25\cm along the beam direction and 0.2\cm in the plane transverse to the beam direction. The primary vertex (PV) is taken to be the vertex corresponding to the hardest scattering in the event, evaluated using tracking information alone, as described in Section 9.4.1 of Ref.~\cite{CMS-TDR-15-02}. In order to select inelastic hadronic collisions, the $\pPb$ events are also required to have at least one tower in each of the HF detectors with energy deposits of more than 3\GeV per tower. Any potential event containing more than one $\pPb$ interaction (pileup) is removed by the offline requirement of a common vertex for the four selected muons, as described below.

\section{Event reconstruction and efficiency}
\label{sec:selections}

Muons are reconstructed by combining information from the silicon tracker and the muon system composed of four layers of detectors (muon ``stations'')~\cite{Sirunyan:2018fpa}. The matching between tracks reconstructed in each of the subsystems proceeds either outside-in, starting from a track in the muon system, or inside-out, starting from a track provided by the silicon tracker. In the latter case, charged particles in the tracker that match track segments in only one or two muon stations are also considered in the analysis to collect low-\pt muons that may not have sufficient energy to traverse the entire muon system. The muons are selected from the reconstructed track candidates that match with at least one segment in any muon station in both $x$ and $y$ coordinates. Furthermore, the muon candidates are required to be identified as ``soft", as defined in Ref.~\cite{Sirunyan:2018fpa}. 
Matching muons to tracks measured in the silicon tracker results in a relative transverse momentum resolution, for muons with \pt up to 100\GeV, of 1\% in the barrel and 3\% in the endcaps~\cite{Sirunyan:2018fpa,CMS:2024qgs}.

All muons are required to pass a variable minimum \pt value based on their pseudorapidity, as described in Table~\ref{tab:fiducialregion}. Muons with opposite-sign charge are combined into pairs with invariant masses $2.6 < m_{\PGm\PGm} < 3.6\GeV$ to form a \JPsi candidate. No pair ambiguities (namely, cases where the two \JPsi mesons can be simultaneously reconstructed with two different combinations of the four muons), nor events with more than four muons satisfying the selection criteria, are found in data. The reconstructed \JPsi candidates are required to have $\pt>6.5\GeV$ and $\abs{y}<2.4$. The four muons are fit to a common vertex, and the $\chi^2$ probability of the fit is required to be larger than $0.1\%$~\cite{Fruhwirth:1987fm}. A total of 16 candidate di-\JPsi events pass all selection criteria (Fig.~\ref{fig:fit}). The requirement of the four muons to share the same vertex suppresses contributions from nonprompt \JPsi mesons originating from the decays of \cPqb quark hadrons. All selected events have vertices consistent with the PV within 100\mum and, therefore, the nonprompt single \JPsi meson contribution is expected to be, at most, a few percent of the prompt one within the selected kinematical range~\cite{CMS:2017dju,ATLAS:2011aqv}. This feature further confirms that the \JPsi mesons considered in this analysis are promptly produced.

Simulations based on the \PYTHIA~8.303~\cite{Sjostrand:2014zea} and \EPOS~\cite{Pierog:2013ria} MC event generators are used to determine the overall efficiency of the analysis including the impact of the dimuon triggers, as well as of the single muon and \JPsi meson reconstruction within the fiducial volume defined in Table~\ref{tab:fiducialregion}. The generated prompt \JPsi mesons are produced unpolarized in agrement with LHC data~\cite{CMS:2013gbz,LHCb:2013izl,ALICE:2020iev}. Generated \PYTHIA-8 $\pp\to \JPsi+\PX$ events are embedded into \pPb\ events simulated with \EPOS\ that replicate the hadronic activity from the underlying event. To improve the agreement between the simulated and measured charged particle multiplicities, a weighting procedure based on the number of tracks per event is applied when combining both MC samples. The generated events are then processed through a detailed simulation of the CMS detector based on the \GEANTfour~\cite{Agostinelli:2002hh} package.
The overall reconstruction efficiency is calculated as the ratio of the number of reconstructed events in the MC simulation (after all selection criteria are applied) to the number of generated events (after generator-level filters) within the fiducial phase space of the analysis.

\section{Signal extraction}

\begin{figure*}[htpb!]
\centering
    \includegraphics[width=.33\textwidth]{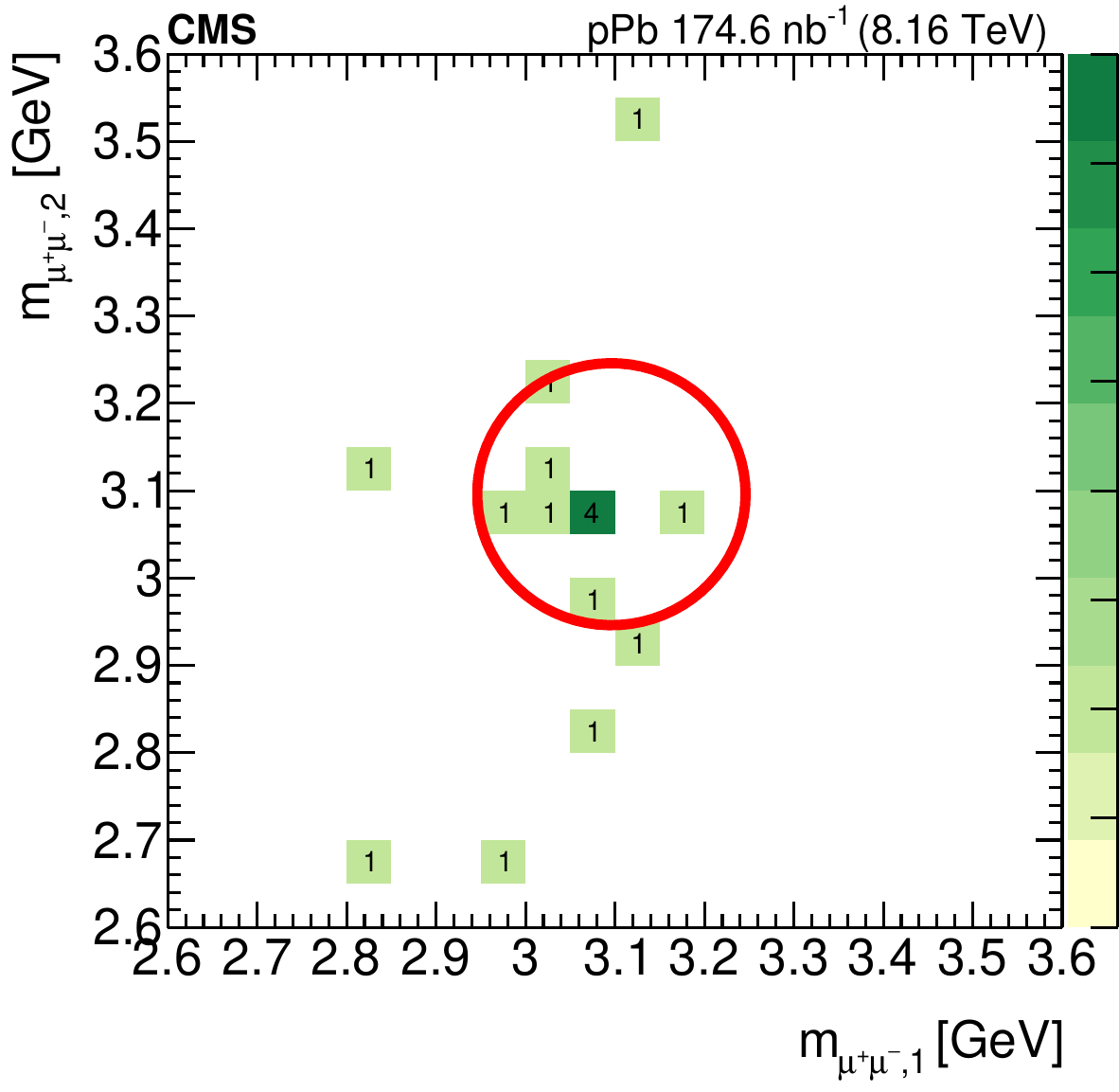}
    \includegraphics[width=.66\textwidth]{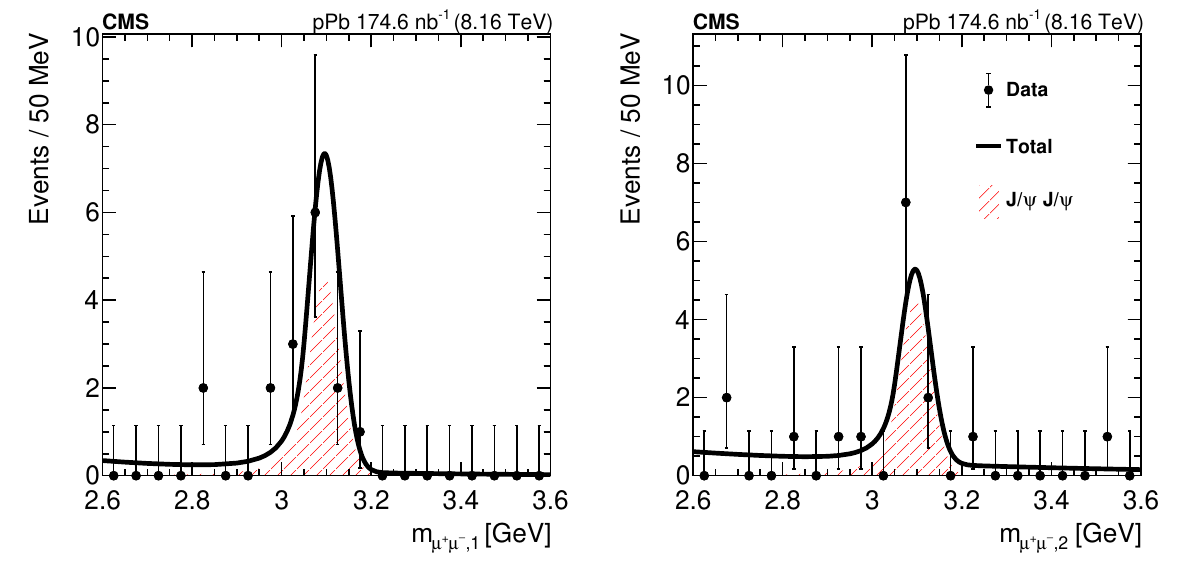}
    \caption{Dimuon invariant mass distributions of the \pPb\ events passing all selection criteria, shown as a 2D map (left panel, with the numbers indicating the counts per bin and the red circle indicating the double \JPsi meson signal region) and its 1D projections for each pair (center and right panels). In the 1D projections, the measured data are represented by the points (with vertical bars showing their Poisson statistical uncertainties), the solid curve shows the overall signal$\,+\,$background fit to the data, and the red hashed area shows the signal yields for the \TJPsi process. \label{fig:fit}
}
\end{figure*}

The signal is extracted from all events passing the fiducial requirements listed in Table~\ref{tab:fiducialregion}, with a two-dimensional (2D) unbinned extended maximum likelihood fit that exploits the invariant masses of the two \JPsi meson candidates as variables. The signal \JPsi mass distribution is modeled with a Crystal-Ball function~\cite{Oreglia:1980cs} and the combinatorial background is described by an exponential function. Both probability density functions (pdfs) are used twice, for the \jpsio and \jpsit states, per event, where the indices 1 and 2 indicate, respectively, the leading (\ie, highest \pt) and subleading \JPsi meson. The parameters describing the Crystal-Ball \JPsi signal model are fixed (for both \JPsi mesons) to values obtained from fits of the simulated distributions, the mean being essentially identical to the world-average \JPsi mass, $m_{\JPsi} = 3.1\GeV$~\cite{ParticleDataGroup:2022pth}, and the relative dimuon mass resolution being close to 1\% ($\delta m_{\JPsi}\approx 30\MeV$). The parameters of the exponential background are left free in the fit.
The per-event 2D map of the two measured dimuon invariant mass distributions is shown in Fig.~\ref{fig:fit} (left) with the red circle indicating the double-\JPsi signal region within $\pm 5 \delta m_{\JPsi}$ around the two \JPsi meson masses.

\begin{table}[htbp!]
  \centering
  \topcaption{Signal and background yields obtained from the likelihood fit procedure in the four-muon and dimuon-dielectron analyses over the $m_{\PGm\PGm,\Pe\Pe}=2.6\mbox{--}3.6\GeV$ mass range.}
  \begin{scotch}{lcc} 
  Yield  &   \multicolumn{2}{c}{Final state}    \\ \hline
                 &  $\mumu\mumu$ & $\mumu\EE$    \\ \hline
  $N(\jpsio^\text{sig}, \jpsit^\text{sig})$ &  $8.5 \pm 3.4$ & $5.7\pm 4.0$  \\
  $N(\jpsio^\text{bkg}, \jpsit^\text{sig})$ &  $0.9 \pm 1.6$ & $4.9\pm 3.8$  \\ 
  $N(\jpsio^\text{sig}, \jpsit^\text{bkg})$ &  $5.2 \pm 3.0$ & $4.5\pm 3.4$  \\ 
  $N(\jpsio^\text{bkg}, \jpsit^\text{bkg})$ &  $1.4 \pm 1.9$ & $6.0\pm 3.5$  \\
  \end{scotch}
\label{tab:fitparameters}
\end{table}

In total, four parameters representing the four combination of yields, $N(\jpsio^\text{sig,bkg}, \jpsit^\text{sig,bkg})$ for each of the two \JPsi meson candidates to be either signal or background, are incorporated into the likelihood fit. The number of signal and background events derived from the fit procedure are listed in Table~\ref{tab:fitparameters}.

The projection of the result of the 2D fit in the two variables, $m_{\PGm\PGm,1}$ and $m_{\PGm\PGm,2}$, is shown in Fig.~\ref{fig:fit} (center and right panels). The data points indicate the measured counts, and the red hashed area shows the fitted signal component. The derived signal yield is $N_\text{sig} \equiv N(\jpsio^\text{sig}, \jpsit^\text{sig}) = \yield$ events. The difference between the red area and the fit (solid black curve) indicates the size of the backgrounds, listed in Table~\ref{tab:fitparameters}, from events with four uncorrelated muons (mostly filling the area to the left of the \JPsi peaks) and with two uncorrelated muons coming from the $\JPsi_{1,2}+\mumu$ combination (filling the empty area below the peaks).

The signal has a significance of 4.9~standard deviations, obtained from the likelihood ratio of the default signal-plus-background fit and the background-only fit (imposing $N_\text{sig}=0$), using the standard asymptotic formula~\cite{Cowan:2010js} assuming that the Wilks theorem conditions apply~\cite{Wilks:1938dza}. As discussed below, the systematic uncertainties are small (6.1\%, with a fraction of them impacting equally signal and background) and do not alter the statistical significance evaluation. This significance was also validated with MC pseudo-experiments.

To fully confirm the observation of the double-\JPsi process, events are also analyzed where the leading \JPsi meson is reconstructed in its $\mumu$ decay mode and the other decays via \jpsiee. This channel has a larger background than the 4-muon one, but increases the total size of the data sample and provides a cross-check of the latter. The electron momentum is estimated by combining the ECAL energy measurement (including the energy sum of all bremsstrahlung photons spatially compatible with originating from the electron trajectory) with the momentum measurement in the tracker. The momentum resolution is typically smaller than 5\% for electrons in the range $2 < \pt < 10\GeV$. It is generally better in the barrel region than in the endcaps, and also depends on the fraction of the electron energy emitted through bremsstrahlung ($f_\text{brem}$) as it traverses the material in front of the ECAL~\cite{CMS:2020uim,CMS-DP-2020-021}. Both electrons are required to have $f_\text{brem}>0.01$ (to remove any potential charged-pion contamination) and to be of opposite-sign charge.

For the $\mumu+\EE$ channel, the selection of the dimuon candidates remains the same as for the double-\mumu analysis, and the candidate $\Pepm$ are required to have $\pt > 2.5\GeV$ and $\abs{\eta}<2.5$ and to share the same common vertex. The parameters of the Crystal-Ball \JPsi meson signal are derived from the MC simulation, with the resulting mass resolution of the dielectron decay mode being $\delta m_{\JPsi}\approx 70\MeV$. A total number of 21 events pass the $(\mumu,\EE)$ criteria with both dileptons having invariant masses over $m_{\PGm\PGm,\Pe\Pe}=2.6\mbox{--}3.6\GeV$. No events passing the selection criteria of the dimuon-dielectron channel are present in the double-dimuon sample.

\begin{figure*}[htpb!]
\centering
    \includegraphics[width=.33\textwidth]{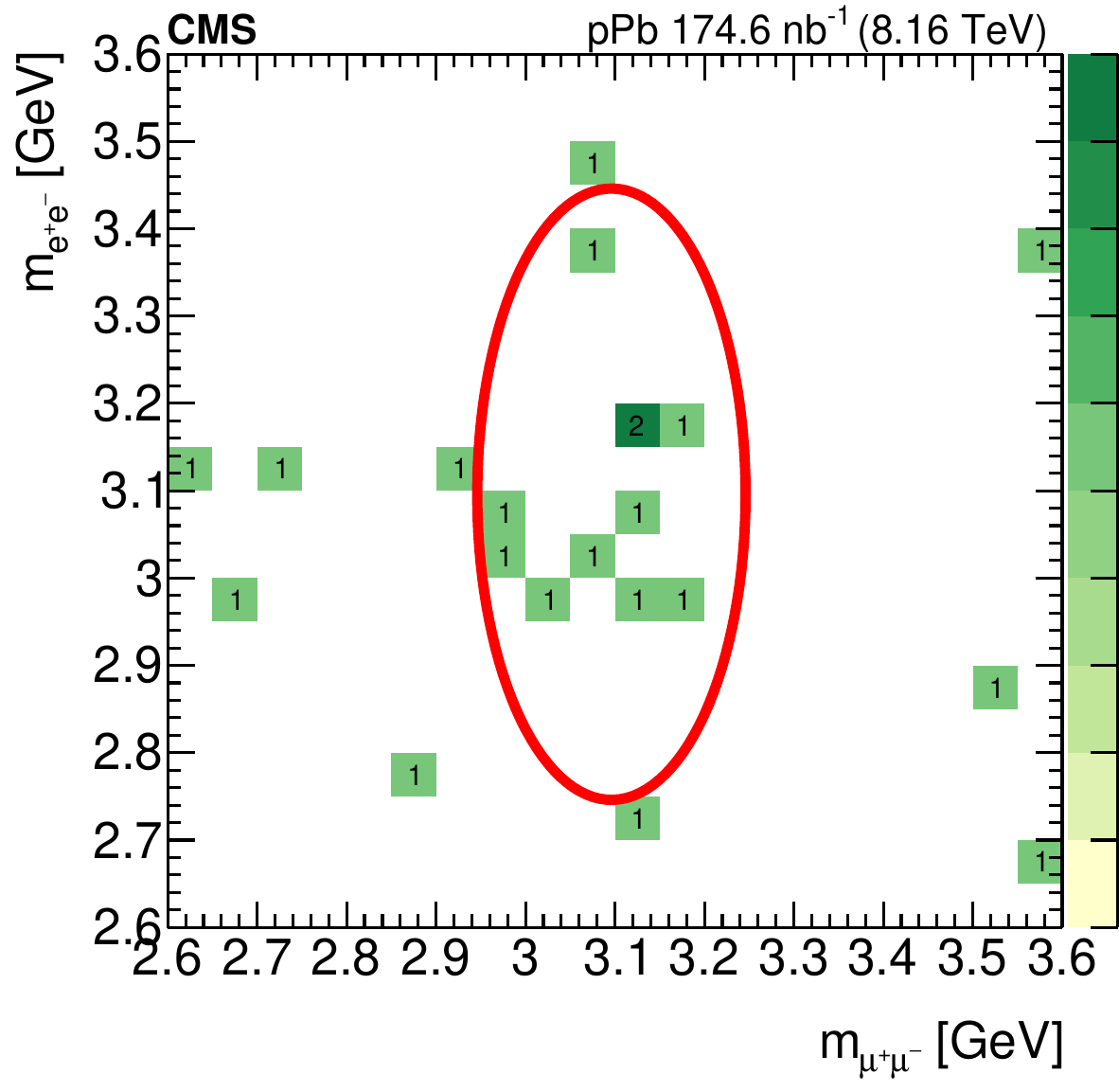}
    \includegraphics[width=.66\textwidth]{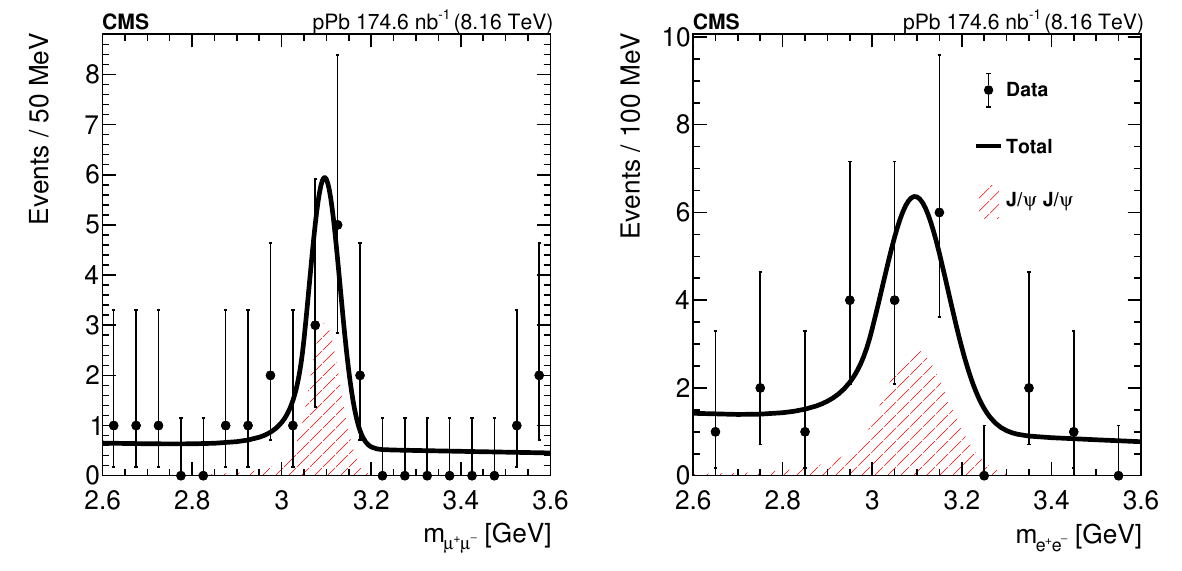}
    \caption{Dilepton invariant mass distributions of the \pPb\ events passing all selection criteria, shown as a 2D map (left panel, with the numbers indicating the counts per bin and the red circle indicating the double \JPsi meson signal region) and its 1D projections for each pair (center and right panels). In the 1D projections, the measured data are represented by the points (vertical bars showing their Poisson statistical uncertainties), the solid curve shows the overall signal$\,+\,$background fit to the data, and the red hashed area shows the signal yields for the \TJPsi process.\label{fig:fit1dprojectionee}
}
\end{figure*}

The fit procedure followed is the same as for the four-muon final state. The per-event 2D map of the dimuon-dielectron invariant mass distributions is shown in Fig.~\ref{fig:fit1dprojectionee} (left) with the red circle indicating the double-\JPsi signal region within $\pm 5 \delta m_{\JPsi}$ around the masses of the two \JPsi mesons. The projection of the result of the 2D fit in the two variables, $m_{\PGm\PGm,1}$ and $m_{\Pe\Pe,2}$, is shown in Fig.~\ref{fig:fit1dprojectionee} (center and right panels). The signal yield (red hashed area) is $5.7\pm 4.0$ events.

The significance of the $(\jpsiee + \jpsimumu)$ mode, calculated in the same way as for the $(\jpsimumu+\jpsimumu)$ channel, is 2.3~standard deviations. The combination of the two decay modes, using the Fisher formalism~\cite{679b84c4-7042-35e5-aa82-ed196e957f94} with $p^\text{value}=p^\text{value}_1p^\text{value}_2\left(1-\log(p^\text{value}_1p^\text{value}_2)\right)$, where $p_{1,2}$ are the $p$-values of each independent analysis, gives a total statistical significance of 5.3~standard deviations.

\section{Results}

The cross section of the double-\JPsi production reconstructed in the four-muon decay mode is obtained through the expression
\begin{equation}
    \xsecbrfull,
\label{eq:sigma_3jpsi_meas}
\end{equation}
where $N_\text{sig} = \yield$ is the number of signal events reconstructed with an overall efficiency $\epsilon$ in the fiducial region defined in Table~\ref{tab:fiducialregion}, $\mathcal{L} = 174.6 \pm 6.1\nbinv$ is the total integrated luminosity, and $\mathcal{B}_{\JPsi\to\PGmp\PGmm} = (5.961 \pm 0.033)\%$ is the dimuon decay branching fraction~\cite{ParticleDataGroup:2022pth}. We do not compute the cross section corresponding to the $(\jpsiee + \jpsimumu)$ decay mode because it has much larger backgrounds and is not precise enough to quantitatively improve the overall cross section extraction. To determine the number of signal events corrected for the reconstruction efficiency, we use the expression $\ntjpsisig/\epsilon = \sum_{i}\ntjpsisig^{i}/\epsilon^{i}$, where $N^i_\text{sig}$ are statistical per-event weights (having a value between 1 and 0, indicating events being more or less signal-like) obtained using the \textsc{SPlot} technique~\cite{Pivk:2004ty}. Based on the kinematics of each \JPsi meson in the \pt and rapidity plane, the $\epsilon^{i}$ values are calculated from a 2D efficiency map obtained from the simulated signal. At central rapidities ($\abs{y}<1$), the efficiency is approximately constant at $\approx90\%$, decreasing towards higher rapidities. The \pt dependence of $\epsilon^{i}$ features a fast rise at low \pt, becoming unity at $\pt\approx 15\GeV$. Any mismatches on the reconstruction and trigger efficiency observed between the MC simulation and the measured data are corrected by applying the scale factors described in Ref.~\cite{Sirunyan:2018fpa}. The double-\JPsi efficiency is $\epsilon=62.1\%$, consistent with the squared single-\JPsi efficiency, as expected, given that it is driven by the kinematics of the two individual \JPsi mesons. 

The systematic uncertainties in the cross section determined with Eq.~(\ref{eq:sigma_3jpsi_meas}) are listed in Table~\ref{tab:systematics}. The dependence of the measured cross section on the modeling of the \JPsi meson signal and background subtraction is examined by replacing the Crystal-Ball pdf of the nominal model by a Crystal-Ball--plus-Gaussian function, as well as the exponential background by a first-order polynomial distribution. The differences observed in the signal yields are 4.0 and 2.5\%, for the variations of the signal and background pdfs, respectively. Releasing the constraint of the \JPsi signal mean and width parameters from the MC simulation results in a negligible difference (0.1\%) and is not considered as a source of systematic uncertainty. A 3.5\% uncertainty is added from the integrated luminosity measurement~\cite{CMS:2021xjt,CMS-PAS-LUM-18-002}. The uncertainty in the scale factors applied to the MC simulation to correct for possible data/MC differences amounts to 1.3\%. Finally, the uncertainty of $\mathcal{B}_{\JPsi\to\PGmp\PGmm}$ propagates into an uncertainty of 1.1\% in the cross section measurement.

\begin{table}[hbtp!]
  \centering
  \topcaption{Relative contributions to the systematic uncertainty in the $\xsec$ measurement. The last row lists the sum in quadrature of all components.}
  \begin{scotch}{lc}
  Source of uncertainty             &  $\xsec$ \\ \hline
  \JPsi meson signal shape          &  4.0\%   \\ 
  Dimuon continuum background shape &  2.5\%   \\ 
  Integrated luminosity             &  3.5\%   \\ 
  Efficiency scale factors          &  1.3\%   \\ 
  Branching fraction                &  1.1\%   \\ [1ex]  
  Total                             &  6.1\%   \\ 
  \end{scotch}
\label{tab:systematics}
\end{table}

The resulting fiducial cross section is $\sigma(\pPb\to\JPsi\JPsi)=\result$. Such a cross section contains in principle contributions from the SPS and DPS production channels shown in Fig.~\ref{fig:diags}. Namely, the result is $\xsec=\sigma^{\pPb \to \TJPsi}_\mathrm{SPS}+\sigma^{\pPb \to \TJPsi}_\mathrm{DPS}$, and the size of each contribution is estimated from theoretical calculations (matched to existing LHC experimental \JPsi data) and the measured double \JPsi meson differential distributions, as explained in the next section.

\section{Discussion}

In the simplest factorized approach represented by Eq.~(\ref{eq:pocketDPS}), the DPS double \JPsi meson cross section can be derived from the squared single \JPsi meson cross section divided by $\sigmaeff$. The single-\JPsi cross sections in \pPb and \pp collisions are well understood both experimentally and theoretically thanks to multiple measurements and calculations that agree with each other within less than 10\%~\cite{Chapon:2020heu}. We use here the perturbative QCD calculation at next-to-leading order (NLO) accuracy from \helaconia~\cite{Shao:2012iz,Shao:2015vga} using the CT14NLO PDF~\cite{Dulat:2015mca} for the proton and the EPPS16 nPDF for the Pb nucleus~\cite{Eskola:2016oht}, reweighted as discussed in Refs.~\cite{Kusina:2017gkz,Kusina:2020dki}. The predicted SPS cross section  (Table~\ref{tab:sigma_jpsi_theory}, first row) is normalized so as to reproduce the inclusive \JPsi measurements in \pPb\ collisions at $\sqrtsNN = 8.16\TeV$~\cite{Lansberg:2016deg}, and thus it does not have any theoretical scale uncertainties assigned but only a ${\approx}6\%$ data normalization uncertainty. The proton-nucleon ($\Pp\mathrm{N}$) cross section (obtained by dividing the \pPb\ cross section by $A = 208$) within the fiducial requirements of Table~\ref{tab:fiducialregion} amounts to $\sigma^{\pPb\to \JPsi+\PX}_\mathrm{SPS}\mathcal{B}(\jpsimumu)/A=21.7\pm 1.4\,\mathrm{(nPDF)}\pm 1.4\,\mathrm{(norm)} \unit{nb}$, to be compared with the corresponding \pp cross section of $\sigma^{\pp\to \JPsi+\PX}_\mathrm{SPS}\mathcal{B}(\jpsimumu)=25.0\pm 1.8\,\mathrm{(PDF)} \unit{nb}$, indicating a ${\approx}10\%$ reduction due to nPDF shadowing. The theoretical prediction for the prompt double \JPsi meson SPS cross section (Table~\ref{tab:sigma_jpsi_theory}, second row) is obtained using the color singlet model at approximately NLO accuracy~\cite{Lansberg:2013qka,Lansberg:2014swa} computed with \helaconia using the CT14NLO proton PDF and the reweighted EPPS16 nPDF for the Pb nucleus. The double \JPsi meson SPS cross section in $\Pp\mathrm{N}$ collisions within the fiducial phase space of Table~\ref{tab:fiducialregion} amounts to $\sigma^{\pPb \to \TJPsi}_\mathrm{SPS}\mathcal{B}^{2}(\jpsimumu)/A=0.097^{+0.185}_{-0.063}\mathrm{(scale)} \pm 0.004\,\mathrm{(nPDF)} \unit{pb}$, to be compared with a \pp cross section of $\sigma^{\pp \to \TJPsi}_\mathrm{SPS}\mathcal{B}^{2}(\jpsimumu)=0.105^{+0.200}_{-0.068}\mathrm{(scale)} \pm 0.004\,\mathrm{(PDF)} \unit{pb}$, indicating about 8\% nPDF shadowing corrections. 

\begin{table}[htpb!]
\centering
\topcaption{Predictions for single and double \JPsi meson production cross sections in SPS processes in \pPb\ collisions at $\sqrtsNN = 8.16\TeV$ within the fiducial phase space of Table~\ref{tab:fiducialregion}, obtained with the \helaconia code using the CT14NLO proton PDF and the reweighted EPPS16 lead nPDF, as explained in the text. The quoted uncertainty of the single (double) \JPsi meson cross section includes nPDF and normalization (nPDF and scale) uncertainties, added in quadrature.
\label{tab:sigma_jpsi_theory}}
\begin{scotch}{lc} 
Process & Theoretical cross section \\ \hline
$\sigma^{\pPb\to\JPsi+\PX}_\mathrm{SPS}\mathcal{B}(\jpsimumu)$ & $4.51\pm 0.42 \unit{$\mu$b}$   \\  [0.5ex]  
$\sigma^{\pPb\to\TJPsi}_\mathrm{SPS}\mathcal{B}^{2}(\jpsimumu)$ & $20.2^{+38.5}_{-13.1}\unit{pb}$  \\  [0.5ex] 
\end{scotch}
\end{table}

\begin{figure}[htpb!]
\centering
\includegraphics[width=.49\textwidth]{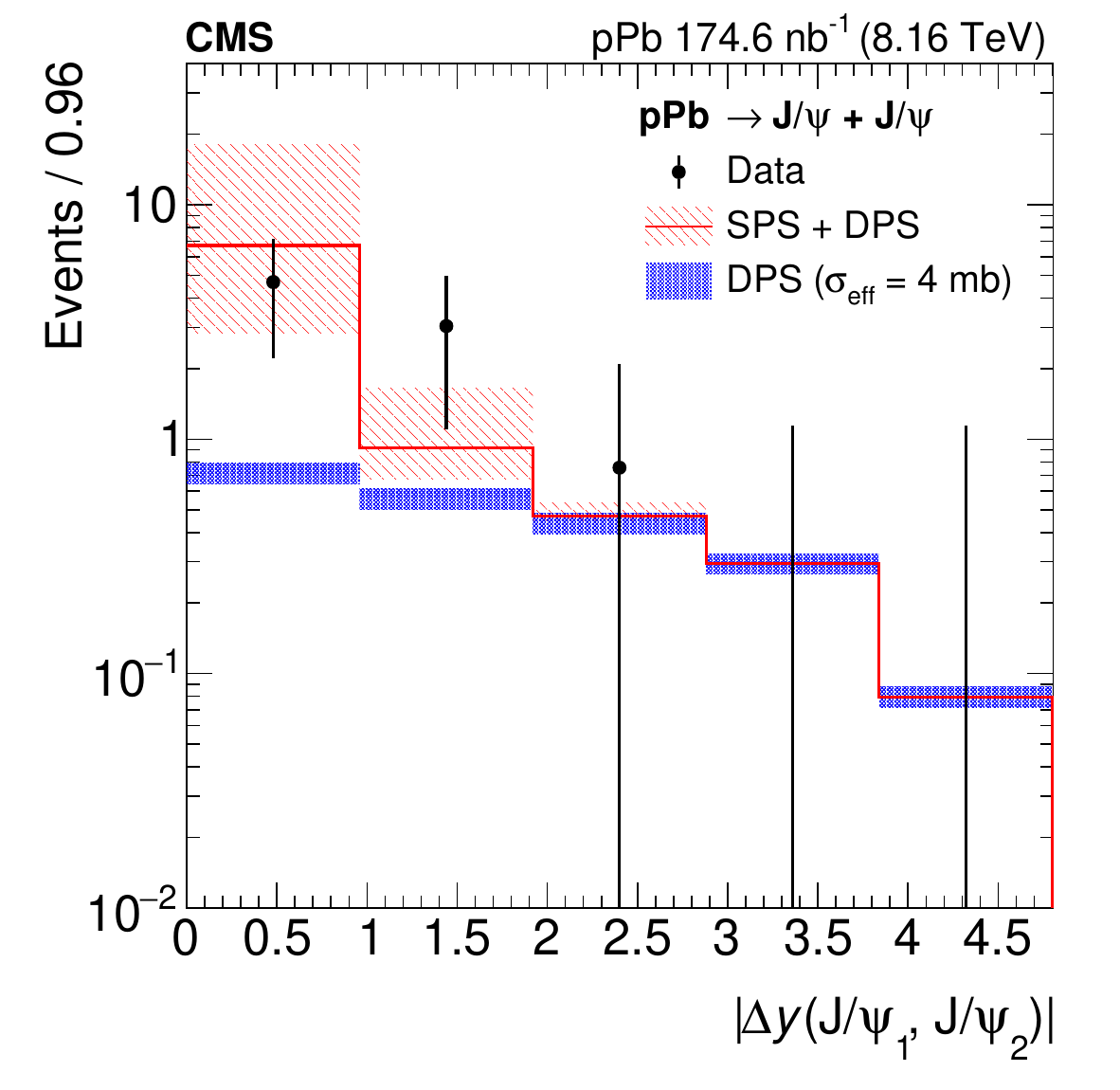}
  \includegraphics[width=.49\textwidth]{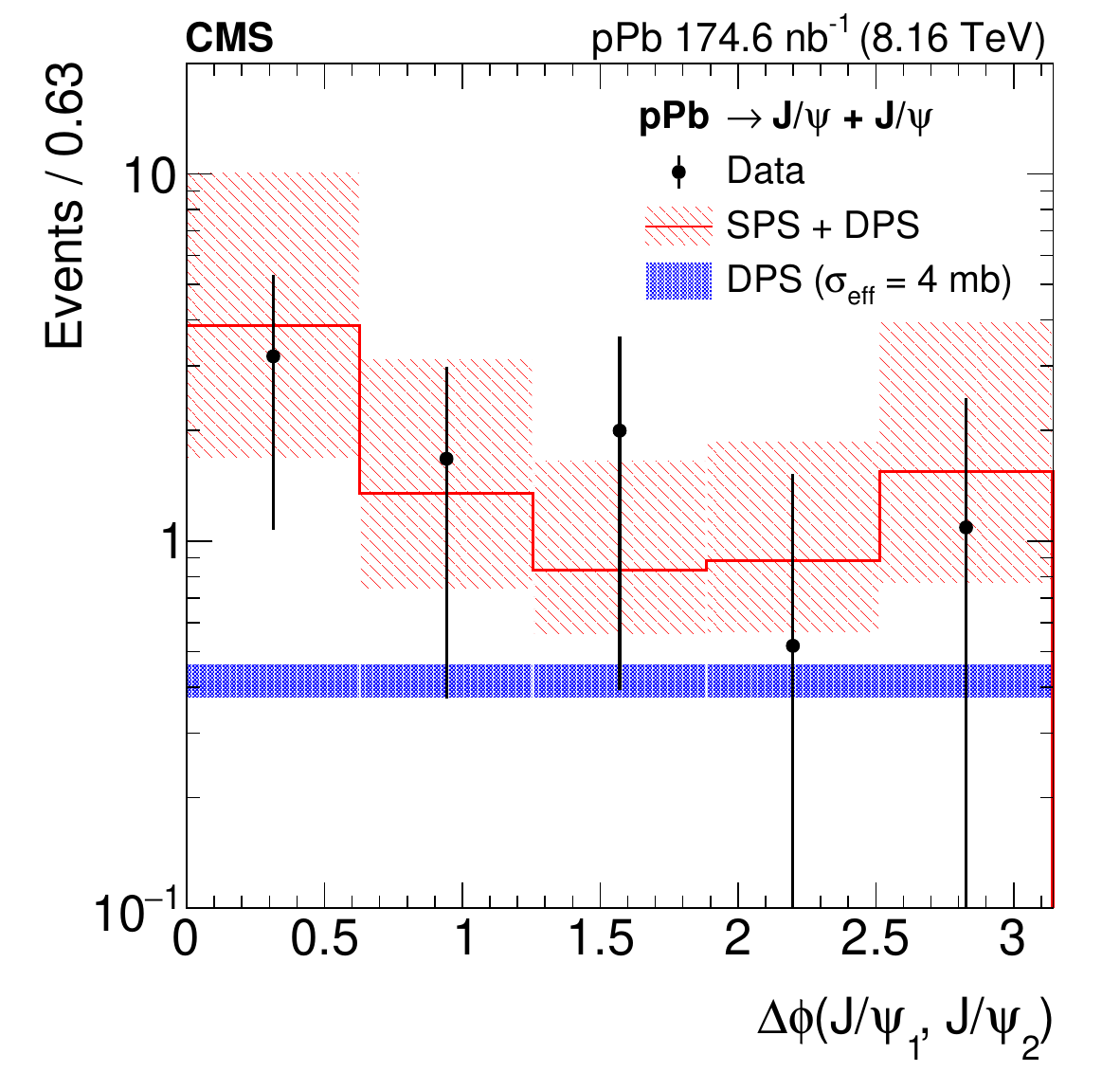}

    \caption{Distribution of rapidity separation (\cmsLeft)  between the two \JPsi mesons in data (black markers), fitted template of the expected DPS contribution (blue histogram) normalized to the data outside of the SPS-dominated region (as described in the text), and fitted SPS$\,+\,$DPS distribution (red histogram). Distribution of azimuthal separation (\cmsRight) between the two \JPsi mesons measured in data (black markers) and expected DPS contribution (blue histogram), and SPS$\,+\,$DPS sum (red histogram), with the relative normalization derived from the \DeltaY distribution (left panel), as explained in the text.
 \label{fig:deltaydeltaphi}
}
\end{figure}

The theoretical SPS single \JPsi meson cross section has an overall uncertainty of ${\pm}9\%$, and thus so will have the corresponding DPS cross section (for a given $\sigmaeff$). However, the SPS double \JPsi meson cross section is poorly constrained, and can vary within a $+200\%,-65\%$ range~\cite{Lansberg:2014swa}. Therefore,  we use our own data in order to constrain the SPS and DPS contributions to the measured double-\JPsi cross section. The most discriminating variables to distinguish SPS from DPS processes are the separation in azimuth \DeltaPhi and rapidity \DeltaY between the two produced particles~\cite{Bartalini:2018qje}. The \JPsi mesons produced by DPS processes are expected to be completely decorrelated in their production and, thus, should feature flatter distributions in their relative azimuthal and rapidity separations (blue histograms in Fig.~\ref{fig:deltaydeltaphi}). On the other hand, the SPS mechanism leads to a correlated production of both \JPsi mesons that prefer to be produced closer in rapidity and with a small or a maximum separation in azimuthal angle. Theoretical expectations based on contributions from multiple SPS channels, computed up to $\mathcal{O}(\alpS^6)$ perturbative accuracy and cross-checked with existing data~\cite{Lansberg:2014swa}, indicate that the distributions of the two \JPsi mesons from SPS processes tend to have a rapidity separation peaked at $\DeltaY = 0$ and maxima at $\DeltaPhi\approx 0,\pi$ (Fig.~\ref{fig:deltaydeltaphi}). 

Due to the limited size of our signal data sample, it is not possible to perform a 2D template fit of the measured $\DeltaPhi\mbox{--}\DeltaY$ distributions with shapes dictated by the theoretical expectations to extract the absolute normalization of the poorly known SPS double \JPsi meson yields. Instead, we identify a phase space region where the SPS contribution is expected to be minimal, and constrain there the expected size of the DPS yield. For this purpose, we use the predicted shape of the \DeltaY distribution, which shows that the $\DeltaY\gtrsim 2$ region is free from SPS contributions (red histogram in Fig.~\ref{fig:deltaydeltaphi}, left) and, therefore, any signal count measured there is dominated by DPS production.

To extract the number of DPS events, we perform a 1D fit to the \DeltaY variable. Using a DPS template built by combining \JPsi mesons from different events, we fit the weighted data in the DPS dominated region of $\DeltaY>2$. The fit results in a yield of 2.1 DPS events and $\sigmaeffdps = 4.0\unit{mb}$, as explained in more detail below, shown as a blue histogram in Fig.~\ref{fig:deltaydeltaphi}. 
The left plot shows the fit result of the \DeltaY distribution compared with the data (markers). We also overlay the combined SPS$\,+\,$DPS distribution shown with the red histogram (and hashed uncertainty bands in red). Figure~\ref{fig:deltaydeltaphi} (right) shows the corresponding \DeltaPhi distribution, with the DPS and SPS$\,+\,$DPS distributions having the same normalization as in the left plot.

This fit procedure allows the extraction of the number of SPS and DPS signal counts: $N_\text{SPS} = 6.4\pm 4.2$ and $N_\text{DPS} =  2.1\pm 2.4$, respectively. The corresponding SPS and DPS fiducial cross sections, derived using Eq.~(\ref{eq:sigma_3jpsi_meas}), amount to $\sigma^{\pPb\to\TJPsi}_\text{SPS}=\resultSPS$ and $\sigma^{\pPb\to\TJPsi}_\text{DPS}=\resultDPS$. From this latter value, we can extract the effective \sigmaeffdpspPb cross section via Eq.~(\ref{eq:pocketDPSpPb}). Using the theoretical predictions (constrained by existing data) for $\sigma^{\pPb \to \JPsi+\PX}_\mathrm{SPS}$ from Table~\ref{tab:sigma_jpsi_theory}, we obtain \sigmaeffdpspPbresult. The extracted $\sigmaeffdpspPb$ value has an arbitrarily large upper uncertainty that indicates that the SPS yield alone would be compatible with the data. 

By neglecting parton correlations, and assuming that the double PDF of the nucleons can be factorized in longitudinal and transverse components, the effective DPS cross section for \pPb collisions can be converted into a \pp-equivalent $\sigmaeff$ one, through Eq.~(\ref{eq:sigmaeffdpspA}). Using $A = 208$ for a Pb nucleus and a factor $F_{\pPb} = 29.5\mbinv$ derived from the \pPb\ thickness function with a Glauber MC model, a value $\sigmaeffdps = 4.0^{+\infty}_{-1.5} \unit{mb}$ is obtained, where the unbound upper uncertainty indicates that the double \JPsi meson yields could in principle be produced without DPS contributions. Therefore, the final result is better expressed as a limit on the maximum amount of DPS allowed by the data, which translates into a lower limit on the effective DPS cross section of $\sigmaeffdps > 1.0 \unit{mb}$ at 95\% confidence level (CL).

\begin{figure*}[htpb!]
\centering
    \includegraphics[width=0.9\textwidth]{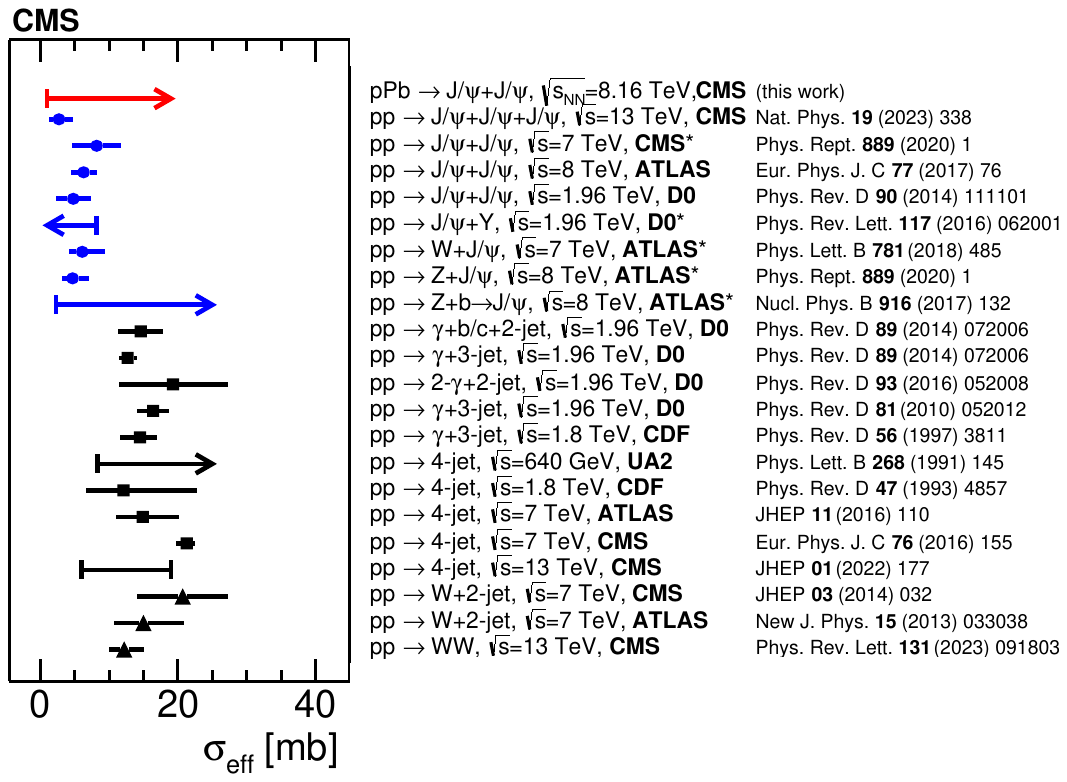}
    \caption{Comparison of the effective DPS cross sections $\sigmaeffdps$ extracted in this work (red arrow lower limit) with the same parameter derived in \pp measurements of double- and triple-quarkonium production~\cite{Lansberg:2019adr,Aaboud:2016fzt,Abazov:2014qba,Shao:2016wor,Aaij:2016bqq,Lansberg:2016muq,Lansberg:2017chq,CMS:2021qsn} (blue circles), as well as in final states with jets~\cite{UA2:1991apc,CDF:1993sbj,D0:2015rpo,Khachatryan:2015pea}, $\gamma+$\,jets~\cite{CDF:1997yfa,D0:2009apj,D0:2014owy}, $\PW+$\,jets~\cite{ATLAS:2013aph,CMS:2013huw}, and double \PW bosons~\cite{CMS:2022pio} (black squares). The asterisk shown in a few legend entries indicates that the result has been obtained by more recent phenomenological analyses of the experimental data.
\label{fig:sigma_eff}}
\end{figure*}

In Fig.~\ref{fig:sigma_eff}, the lower $\sigmaeffdps$ bound determined in this work is compared with other extractions of the effective DPS cross section derived from double- and triple-quarkonium production measurements~\cite{Lansberg:2017chq,Lansberg:2019adr,ATLAS:2014ofp,Abazov:2015fbl,Aaij:2016bqq,Abazov:2014qba,Aaboud:2016fzt,Aaij:2016bqq,Abazov:2014qba,Aaboud:2016fzt,Shao:2016wor,Lansberg:2016muq,CMS:2021qsn} (blue circles), as well as from processes with jets, photons, and $\PW$ bosons~\cite{UA2:1991apc,CDF:1993sbj,CDF:1997yfa,D0:2009apj,ATLAS:2013aph,CMS:2022pio,Khachatryan:2015pea} (black squares). A few of the quarkonium $\sigmaeffdps$ values plotted (indicated with an asterisk in the legend) are those derived by more recent phenomenological studies~\cite{Lansberg:2016muq,Shao:2016wor,Lansberg:2017chq,Lansberg:2019adr} of the experimental data, with an improved evaluation of the SPS contributions. Overall, the effective cross sections obtained from multi-quarkonium production favor a smaller value of $\sigmaeffdps \approx 3\mbox{--}10\unit{mb}$ compared with the $\sigmaeffdps \approx 10\mbox{--}20\unit{mb}$ result of the other types of multiple hard scattering processes. Our measurement of $\sigmaeffdps > 1.0 \unit{mb}$ at 95\% CL calls for larger \pPb\ data sets to confirm, or not, its consistency with the $\sigmaeff$ values derived from \pp data and, eventually, better constrain the average effective interparton distance squared ($\sigmaeff$) involved in the production of a pair of quarkonium mesons in hadronic collisions.

\section{Summary}

The first observation of the concurrent production of two \JPsi mesons in proton-lead (\pPb) collisions has been reported. Events with two \JPsi mesons, each decaying into two muons, have been reconstructed in \pPb\ collisions at a nucleon-nucleon center-of-mass-energy of $\sqrtsNN = 8.16\TeV$. The data sample corresponds to an integrated luminosity of 174.6\nbinv collected by the CMS experiment at the CERN LHC. After all selection requirements, \yield events are found that confirm the production of the $\TJPsi$ final state. Events where the least energetic \JPsi meson is reconstructed in the dielectron channel are also considered in the search, leading to a signal yield of $5.7 \pm 4.0$ events. The statistical significance of the signal, relative to the background-only expectation, corresponds to 5.3 standard deviations (4.9 standard deviations in the four-muon channel alone). The fiducial cross section for $\TJPsi$ production is $\sigma(\pPb\to\TJPsi)= \result$. This result is compared with the theoretical expectations for $\TJPsi$ production via the sum of contributions from single- and double-parton scatterings (SPS and DPS, respectively). Under the simplest assumption of factorization of multiple hard scattering probabilities in terms of SPS cross sections, the measured $\TJPsi$ cross section is consistent with the production of SPS$\,+\,$DPS processes with fiducial cross sections $\sigma^{\pPb\to\TJPsi}_\text{SPS}=\resultSPS$ and $\sigma^{\pPb\to\TJPsi}_\text{DPS}=\resultDPS$. The derived DPS cross section can be transformed into a lower bound on the effective DPS cross section parameter ($\sigmaeffdps$) that is closely related to the squared average interparton transverse separation in the collision. A limit of $\sigmaeffdps > 1.0 \unit{mb}$ at 95\% confidence level is set, which is consistent with $\sigmaeff$ values obtained from double and triple quarkonium measurements in \pp collisions. The present analysis supports the interest of exploiting the production of multiple heavy and/or high-\pt particles in \pPb collisions at the LHC as a novel means to study the dynamics of multiple independent hard scatterings.

\begin{acknowledgments}

We congratulate our colleagues in the CERN accelerator departments for the excellent performance of the LHC and thank the technical and administrative staffs at CERN and at other CMS institutes for their contributions to the success of the CMS effort. In addition, we gratefully acknowledge the computing centers and personnel of the Worldwide LHC Computing Grid and other centers for delivering so effectively the computing infrastructure essential to our analyses. Finally, we acknowledge the enduring support for the construction and operation of the LHC, the CMS detector, and the supporting computing infrastructure provided by the following funding agencies: SC (Armenia), BMBWF and FWF (Austria); FNRS and FWO (Belgium); CNPq, CAPES, FAPERJ, FAPERGS, and FAPESP (Brazil); MES and BNSF (Bulgaria); CERN; CAS, MoST, and NSFC (China); MINCIENCIAS (Colombia); MSES and CSF (Croatia); RIF (Cyprus); SENESCYT (Ecuador); ERC PRG, RVTT3 and MoER TK202 (Estonia); Academy of Finland, MEC, and HIP (Finland); CEA and CNRS/IN2P3 (France); SRNSF (Georgia); BMBF, DFG, and HGF (Germany); GSRI (Greece); NKFIH (Hungary); DAE and DST (India); IPM (Iran); SFI (Ireland); INFN (Italy); MSIP and NRF (Republic of Korea); MES (Latvia); LMTLT (Lithuania); MOE and UM (Malaysia); BUAP, CINVESTAV, CONACYT, LNS, SEP, and UASLP-FAI (Mexico); MOS (Montenegro); MBIE (New Zealand); PAEC (Pakistan); MES and NSC (Poland); FCT (Portugal); MESTD (Serbia); MCIN/AEI and PCTI (Spain); MOSTR (Sri Lanka); Swiss Funding Agencies (Switzerland); MST (Taipei); MHESI and NSTDA (Thailand); TUBITAK and TENMAK (Turkey); NASU (Ukraine); STFC (United Kingdom); DOE and NSF (USA).

\hyphenation{Rachada-pisek} Individuals have received support from the Marie-Curie program and the European Research Council and Horizon 2020 Grant, contract Nos.\ 675440, 724704, 752730, 758316, 765710, 824093, 101115353, 101002207, and COST Action CA16108 (European Union); the Leventis Foundation; the Alfred P.\ Sloan Foundation; the Alexander von Humboldt Foundation; the Science Committee, project no. 22rl-037 (Armenia); the Belgian Federal Science Policy Office; the Fonds pour la Formation \`a la Recherche dans l'Industrie et dans l'Agriculture (FRIA-Belgium); the F.R.S.-FNRS and FWO (Belgium) under the ``Excellence of Science -- EOS" -- be.h project n.\ 30820817; the Beijing Municipal Science \& Technology Commission, No. Z191100007219010 and Fundamental Research Funds for the Central Universities (China); the Ministry of Education, Youth and Sports (MEYS) of the Czech Republic; the Shota Rustaveli National Science Foundation, grant FR-22-985 (Georgia); the Deutsche Forschungsgemeinschaft (DFG), under Germany's Excellence Strategy -- EXC 2121 ``Quantum Universe" -- 390833306, and under project number 400140256 - GRK2497; the Hellenic Foundation for Research and Innovation (HFRI), Project Number 2288 (Greece); the Hungarian Academy of Sciences, the New National Excellence Program - \'UNKP, the NKFIH research grants K 131991, K 133046, K 138136, K 143460, K 143477, K 146913, K 146914, K 147048, 2020-2.2.1-ED-2021-00181, and TKP2021-NKTA-64 (Hungary); the Council of Science and Industrial Research, India; ICSC -- National Research Center for High Performance Computing, Big Data and Quantum Computing and FAIR -- Future Artificial Intelligence Research, funded by the NextGenerationEU program (Italy); the Latvian Council of Science; the Ministry of Education and Science, project no. 2022/WK/14, and the National Science Center, contracts Opus 2021/41/B/ST2/01369 and 2021/43/B/ST2/01552 (Poland); the Funda\c{c}\~ao para a Ci\^encia e a Tecnologia, grant CEECIND/01334/2018 (Portugal); the National Priorities Research Program by Qatar National Research Fund; MCIN/AEI/10.13039/501100011033, ERDF ``a way of making Europe", and the Programa Estatal de Fomento de la Investigaci{\'o}n Cient{\'i}fica y T{\'e}cnica de Excelencia Mar\'{\i}a de Maeztu, grant MDM-2017-0765 and Programa Severo Ochoa del Principado de Asturias (Spain); the Chulalongkorn Academic into Its 2nd Century Project Advancement Project, and the National Science, Research and Innovation Fund via the Program Management Unit for Human Resources \& Institutional Development, Research and Innovation, grant B37G660013 (Thailand); the Kavli Foundation; the Nvidia Corporation; the SuperMicro Corporation; the Welch Foundation, contract C-1845; and the Weston Havens Foundation (USA).
\end{acknowledgments}

\bibliography{auto_generated}

\providecommand{\href}[2]{#2}\begingroup\raggedright\begin{thebibliography}{10}%
\makeatletter
\providecommand{\hrefCMSnoop }[0]{\@secondoftwo}%
\makeatother
\providecommand{\doi}{\texttt{doi:}\begingroup \urlstyle{tt}\Url}

\bibitem{Diehl:2017wew}
\hrefCMSnoop {}{M.~Diehl and J.~R. Gaunt, ``Double parton scattering theory
  overview'',} \textit{ Adv. Ser. Direct. High Energy Phys.} \textbf{ 29}
  (2018) 7,
  \href{http://dx.doi.org/10.1142/9789813227767_0002}{\doi{10.1142/9789813227767_0002}},
  \href{http://www.arXiv.org/abs/1710.04408}{\texttt{arXiv:1710.04408}}.

\bibitem{Bartalini:2018qje}
P.~Bartalini and J.~R. Gaunt, eds., ``Multiple parton interactions at the
  {LHC}'', volume~29.
\newblock World Scientific, 2019.
\newblock \href{http://dx.doi.org/10.1142/10646}{\doi{10.1142/10646}},
  ISBN~978-981-322-775-0, 978-981-322-777-4.

\bibitem{CMS:2019wch}
\hrefCMSnoop {}{{CMS Collaboration}, ``Search for {Higgs} and {Z} boson decays
  to {J}/\ensuremath{\psi} or {Y} pairs in the four-muon final state in
  proton-proton collisions at $\sqrt{s} =13$~{TeV}'',} \textit{ Phys. Lett. B}
  \textbf{ 797} (2019) 134811,
  \href{http://dx.doi.org/10.1016/j.physletb.2019.134811}{\doi{10.1016/j.physletb.2019.134811}},
  \href{http://www.arXiv.org/abs/1905.10408}{\texttt{arXiv:1905.10408}}.

\bibitem{Sirunyan:2020txn}
\hrefCMSnoop {}{{CMS Collaboration}, ``Measurement of the {$\Upsilon$(1S)} pair
  production cross section and search for resonances decaying to
  {$\Upsilon$(1S)$\mu^+\mu^-$} in proton-proton collisions at $\sqrt{s} =
  13$~{TeV}'',} \textit{ Phys. Lett. B} \textbf{ 808} (2020) 135578,
  \href{http://dx.doi.org/10.1016/j.physletb.2020.135578}{\doi{10.1016/j.physletb.2020.135578}},
  \href{http://www.arXiv.org/abs/2002.06393}{\texttt{arXiv:2002.06393}}.

\bibitem{CMS:2023owd}
\hrefCMSnoop {}{{CMS Collaboration}, ``New structures in the
  {J/\ensuremath{\psi}J/\ensuremath{\psi}} mass spectrum in proton-proton
  collisions at $\sqrt{s} = 13$~{TeV}'',} \textit{ Phys. Rev. Lett.} \textbf{
  132} (2024) 111901,
  \href{http://dx.doi.org/10.1103/PhysRevLett.132.111901}{\doi{10.1103/PhysRevLett.132.111901}},
  \href{http://www.arXiv.org/abs/2306.07164}{\texttt{arXiv:2306.07164}}.

\bibitem{dEnterria:2017yhd}
\hrefCMSnoop {}{D.~d'Enterria and A.~Snigirev, ``Double, triple, and $n$-parton
  scatterings in high-energy proton and nuclear collisions'',} \textit{ Adv.
  Ser. Direct. High Energy Phys.} \textbf{ 29} (2018) 159,
  \href{http://dx.doi.org/10.1142/9789813227767_0009}{\doi{10.1142/9789813227767_0009}},
  \href{http://www.arXiv.org/abs/1708.07519}{\texttt{arXiv:1708.07519}}.

\bibitem{Chapon:2020heu}
\hrefCMSnoop {}{E.~Chapon { et~al.}, ``Prospects for quarkonium studies at the
  high-luminosity {LHC}'',} \textit{ Prog. Part. Nucl. Phys.} \textbf{ 122}
  (2022) 103906,
  \href{http://dx.doi.org/10.1016/j.ppnp.2021.103906}{\doi{10.1016/j.ppnp.2021.103906}},
  \href{http://www.arXiv.org/abs/2012.14161}{\texttt{arXiv:2012.14161}}.

\bibitem{Sjostrand:2017cdm}
\hrefCMSnoop {}{T.~Sj{\"o}strand, ``The development of {MPI} modeling in
  {PYTHIA}'',} \textit{ Adv. Ser. Direct. High Energy Phys.} \textbf{ 29}
  (2019) 191,
  \href{http://dx.doi.org/10.1142/9789813227767\_0010}{\doi{10.1142/9789813227767\_0010}},
  \href{http://www.arXiv.org/abs/1706.02166}{\texttt{arXiv:1706.02166}}.

\bibitem{Seymour:2013qka}
\hrefCMSnoop {}{M.~H. Seymour and A.~Siodmok, ``Constraining {MPI} models using
  $\sigma_\mathrm{eff}$ and recent {Tevatron} and {LHC} underlying event
  data'',} \textit{ JHEP} \textbf{ 10} (2013) 113,
  \href{http://dx.doi.org/10.1007/JHEP10(2013)113}{\doi{10.1007/JHEP10(2013)113}},
  \href{http://www.arXiv.org/abs/1307.5015}{\texttt{arXiv:1307.5015}}.

\bibitem{Blok:2017alw}
\hrefCMSnoop {}{B.~Blok and M.~Strikman, ``Multiparton pp and {pA} collisions:
  From geometry to parton\textendash{}parton correlations'',} \textit{ Adv.
  Ser. Direct. High Energy Phys.} \textbf{ 29} (2018) 63,
  \href{http://dx.doi.org/10.1142/9789813227767_0005}{\doi{10.1142/9789813227767_0005}},
  \href{http://www.arXiv.org/abs/1709.00334}{\texttt{arXiv:1709.00334}}.

\bibitem{Aaij:2011yc}
\hrefCMSnoop {}{{LHCb Collaboration}, ``Observation of {J}$/\psi$ pair
  production in pp collisions at $\sqrt{s}=7$~{TeV}'',} \textit{ Phys. Lett. B}
  \textbf{ 707} (2012) 52,
  \href{http://dx.doi.org/10.1016/j.physletb.2011.12.015}{\doi{10.1016/j.physletb.2011.12.015}},
\href{http://www.arXiv.org/abs/1109.0963}{\texttt{arXiv:1109.0963}}.

\bibitem{Abazov:2014qba}
\hrefCMSnoop {}{{D0} Collaboration, ``Observation and studies of double
  {J}$/\psi$ production at the {Tevatron}'',} \textit{ Phys. Rev. D} \textbf{
  90} (2014) 111101,
  \href{http://dx.doi.org/10.1103/PhysRevD.90.111101}{\doi{10.1103/PhysRevD.90.111101}},
\href{http://www.arXiv.org/abs/1406.2380}{\texttt{arXiv:1406.2380}}.

\bibitem{Khachatryan:2014iia}
\hrefCMSnoop {}{{CMS Collaboration}, ``Measurement of prompt {J}$/\psi$ pair
  production in pp collisions at $ \sqrt{s}= 7$~{TeV}'',} \textit{ JHEP}
  \textbf{ 09} (2014) 094,
  \href{http://dx.doi.org/10.1007/JHEP09(2014)094}{\doi{10.1007/JHEP09(2014)094}},
\href{http://www.arXiv.org/abs/1406.0484}{\texttt{arXiv:1406.0484}}.

\bibitem{Aaboud:2016fzt}
\hrefCMSnoop {}{{ATLAS Collaboration}, ``Measurement of the prompt {J}$/\psi$
  pair production cross-section in pp collisions at $\sqrt{s} = 8$~{TeV} with
  the {ATLAS} detector'',} \textit{ Eur. Phys. J. C} \textbf{ 77} (2017) 76,
  \href{http://dx.doi.org/10.1140/epjc/s10052-017-4644-9}{\doi{10.1140/epjc/s10052-017-4644-9}},
\href{http://www.arXiv.org/abs/1612.02950}{\texttt{arXiv:1612.02950}}.

\bibitem{Aaij:2016bqq}
\hrefCMSnoop {}{{LHCb Collaboration}, ``Measurement of the {J}$/\psi$ pair
  production cross-section in pp collisions at $\sqrt{s}=13$~{TeV}'',} \textit{
  JHEP} \textbf{ 06} (2017) 047,
  \href{http://dx.doi.org/10.1007/JHEP06(2017)047}{\doi{10.1007/JHEP06(2017)047}},
  \href{http://www.arXiv.org/abs/1612.07451}{\texttt{arXiv:1612.07451}}.
[Erratum: \DOI{10.1007/JHEP10(2017)068}].

\bibitem{CMS:2021qsn}
\hrefCMSnoop {}{{CMS Collaboration}, ``Observation of triple
  {J}/\ensuremath{\psi} meson production in proton-proton collisions'',}
  \textit{ Nature Phys.} \textbf{ 19} (2023) 338,
  \href{http://dx.doi.org/10.1038/s41567-022-01838-y}{\doi{10.1038/s41567-022-01838-y}},
  \href{http://www.arXiv.org/abs/2111.05370}{\texttt{arXiv:2111.05370}}.
  [Erratum: \DOI{10.1038/s41567-023-01992-x}].

\bibitem{Abazov:2015fbl}
\hrefCMSnoop {}{{D0} Collaboration, ``Evidence for simultaneous production of
  {J}$/\psi$ and {$\Upsilon$} mesons'',} \textit{ Phys. Rev. Lett.} \textbf{
  116} (2016) 082002,
  \href{http://dx.doi.org/10.1103/PhysRevLett.116.082002}{\doi{10.1103/PhysRevLett.116.082002}},
\href{http://www.arXiv.org/abs/1511.02428}{\texttt{arXiv:1511.02428}}.

\bibitem{Khachatryan:2016ydm}
\hrefCMSnoop {}{{CMS Collaboration}, ``Observation of {$\Upsilon$(1S)} pair
  production in proton-proton collisions at $ \sqrt{s}=8$~{TeV}'',} \textit{
  JHEP} \textbf{ 05} (2017) 013,
  \href{http://dx.doi.org/10.1007/JHEP05(2017)013}{\doi{10.1007/JHEP05(2017)013}},
\href{http://www.arXiv.org/abs/1610.07095}{\texttt{arXiv:1610.07095}}.

\bibitem{ATLAS:2013aph}
\hrefCMSnoop {}{{ATLAS Collaboration}, ``Measurement of hard double-parton
  interactions in {W}$(\to \ell\nu)$+ 2 jet events at $\sqrt{s}=7$~{TeV} with
  the {ATLAS} detector'',} \textit{ New J. Phys.} \textbf{ 15} (2013) 033038,
  \href{http://dx.doi.org/10.1088/1367-2630/15/3/033038}{\doi{10.1088/1367-2630/15/3/033038}},
  \href{http://www.arXiv.org/abs/1301.6872}{\texttt{arXiv:1301.6872}}.

\bibitem{CMS:2013huw}
\hrefCMSnoop {}{{CMS Collaboration}, ``Study of double parton scattering using
  {W} + 2-jet events in proton-proton collisions at $\sqrt{s} = 7$~{TeV}'',}
  \textit{ JHEP} \textbf{ 03} (2014) 032,
  \href{http://dx.doi.org/10.1007/JHEP03(2014)032}{\doi{10.1007/JHEP03(2014)032}},
  \href{http://www.arXiv.org/abs/1312.5729}{\texttt{arXiv:1312.5729}}.

\bibitem{ATLAS:2016rnd}
\hrefCMSnoop {}{{ATLAS Collaboration}, ``Study of hard double-parton scattering
  in four-jet events in pp collisions at $\sqrt{s}=7$~{TeV} with the {ATLAS}
  experiment'',} \textit{ JHEP} \textbf{ 11} (2016) 110,
  \href{http://dx.doi.org/10.1007/JHEP11(2016)110}{\doi{10.1007/JHEP11(2016)110}},
  \href{http://www.arXiv.org/abs/1608.01857}{\texttt{arXiv:1608.01857}}.

\bibitem{CMS:2017han}
\hrefCMSnoop {}{{CMS Collaboration}, ``Constraints on the double-parton
  scattering cross section from same-sign {W} boson pair production in
  proton-proton collisions at $\sqrt{s}=8$~{TeV}'',} \textit{ JHEP} \textbf{
  02} (2018) 032,
  \href{http://dx.doi.org/10.1007/JHEP02(2018)032}{\doi{10.1007/JHEP02(2018)032}},
  \href{http://www.arXiv.org/abs/1712.02280}{\texttt{arXiv:1712.02280}}.

\bibitem{Aaboud:2018tiq}
\hrefCMSnoop {}{{ATLAS Collaboration}, ``Study of the hard double-parton
  scattering contribution to inclusive four-lepton production in pp collisions
  at $\sqrt{s} =8$~{TeV} with the {ATLAS} detector'',} \textit{ Phys. Lett. B}
  \textbf{ 790} (2019) 595,
  \href{http://dx.doi.org/10.1016/j.physletb.2019.01.062}{\doi{10.1016/j.physletb.2019.01.062}},
  \href{http://www.arXiv.org/abs/1811.11094}{\texttt{arXiv:1811.11094}}.

\bibitem{CMS:2019jcb}
\hrefCMSnoop {}{{CMS Collaboration}, ``Evidence for {WW} production from
  double-parton interactions in proton\textendash{}proton collisions at
  $\sqrt{s} = 13$~{TeV}'',} \textit{ Eur. Phys. J. C} \textbf{ 80} (2020) 41,
  \href{http://dx.doi.org/10.1140/epjc/s10052-019-7541-6}{\doi{10.1140/epjc/s10052-019-7541-6}},
  \href{http://www.arXiv.org/abs/1909.06265}{\texttt{arXiv:1909.06265}}.

\bibitem{CMS:2021lxi}
\hrefCMSnoop {}{{CMS Collaboration}, ``Measurement of double-parton scattering
  in inclusive production of four jets with low transverse momentum in
  proton-proton collisions at $\sqrt{s} = 13$~{TeV}'',} \textit{ JHEP} \textbf{
  01} (2022) 177,
  \href{http://dx.doi.org/10.1007/JHEP01(2022)177}{\doi{10.1007/JHEP01(2022)177}},
  \href{http://www.arXiv.org/abs/2109.13822}{\texttt{arXiv:2109.13822}}.

\bibitem{CMS:2022pio}
\hrefCMSnoop {}{{CMS Collaboration}, ``{Observation of same-sign WW production
  from double parton scattering in proton-proton collisions at $\sqrt{s} =
  13$~TeV}'',} \textit{ Phys. Rev. Lett.} \textbf{ 131} (2023) 091803,
  \href{http://dx.doi.org/10.1103/PhysRevLett.131.091803}{\doi{10.1103/PhysRevLett.131.091803}},
  \href{http://www.arXiv.org/abs/2206.02681}{\texttt{arXiv:2206.02681}}.

\bibitem{Huayra:2023gio}
\hrefCMSnoop {}{E.~Huayra, J.~V.~C. Lovato, and E.~G. de~Oliveira, ``Valence
  and sea parton correlations in double parton scattering from data'',}
  \textit{ JHEP} \textbf{ 09} (2023) 177,
  \href{http://dx.doi.org/10.1007/JHEP09(2023)177}{\doi{10.1007/JHEP09(2023)177}},
  \href{http://www.arXiv.org/abs/2305.11106}{\texttt{arXiv:2305.11106}}.

\bibitem{Strikman:2001gz}
\hrefCMSnoop {}{M.~Strikman and D.~Treleani, ``Measuring double parton
  distributions in nucleons at proton nucleus colliders'',} \textit{ Phys. Rev.
  Lett.} \textbf{ 88} (2002) 031801,
  \href{http://dx.doi.org/10.1103/PhysRevLett.88.031801}{\doi{10.1103/PhysRevLett.88.031801}},
  \href{http://www.arXiv.org/abs/hep-ph/0111468}{\texttt{arXiv:hep-ph/0111468}}.

\bibitem{Cattaruzza:2004qb}
\hrefCMSnoop {}{E.~Cattaruzza, A.~Del~Fabbro, and D.~Treleani, ``Heavy-quark
  production in proton-nucleus collision at the {CERN LHC}'',} \textit{ Phys.
  Rev. D} \textbf{ 70} (2004) 034022,
  \href{http://dx.doi.org/10.1103/PhysRevD.70.034022}{\doi{10.1103/PhysRevD.70.034022}},
  \href{http://www.arXiv.org/abs/hep-ph/0404177}{\texttt{arXiv:hep-ph/0404177}}.

\bibitem{dEnterria:2012jam}
\hrefCMSnoop {}{D.~d'Enterria and A.~M. Snigirev, ``Same-sign {WW} production
  in proton-nucleus collisions at the {LHC} as a signal for double parton
  scattering'',} \textit{ Phys. Lett. B} \textbf{ 718} (2013) 1395,
  \href{http://dx.doi.org/10.1016/j.physletb.2012.12.032}{\doi{10.1016/j.physletb.2012.12.032}},
  \href{http://www.arXiv.org/abs/1211.0197}{\texttt{arXiv:1211.0197}}.

\bibitem{Shao:2020acd}
\hrefCMSnoop {}{H.-S. Shao, ``Probing impact-parameter dependent nuclear parton
  densities from double parton scatterings in heavy-ion collisions'',} \textit{
  Phys. Rev. D} \textbf{ 101} (2020) 054036,
  \href{http://dx.doi.org/10.1103/PhysRevD.101.054036}{\doi{10.1103/PhysRevD.101.054036}},
  \href{http://www.arXiv.org/abs/2001.04256}{\texttt{arXiv:2001.04256}}.

\bibitem{dEnterria:2020dwq}
\hrefCMSnoop {}{D.~d'Enterria and C.~Loizides, ``Progress in the {Glauber}
  model at collider energies'',} \textit{ Ann. Rev. Nucl. Part. Sci.} \textbf{
  71} (2021) 315,
  \href{http://dx.doi.org/10.1146/annurev-nucl-102419-060007}{\doi{10.1146/annurev-nucl-102419-060007}},
  \href{http://www.arXiv.org/abs/2011.14909}{\texttt{arXiv:2011.14909}}.

\bibitem{CMS:2015nfb}
\hrefCMSnoop {}{{CMS Collaboration}, ``Measurement of the inelastic cross
  section in proton-lead collisions at $\sqrt{{s_{_\mathrm{NN}}}}=
  5.02$~{TeV}'',} \textit{ Phys. Lett. B} \textbf{ 759} (2016) 641,
  \href{http://dx.doi.org/10.1016/j.physletb.2016.06.027}{\doi{10.1016/j.physletb.2016.06.027}},
  \href{http://www.arXiv.org/abs/1509.03893}{\texttt{arXiv:1509.03893}}.

\bibitem{Loizides:2017ack}
\hrefCMSnoop {}{C.~Loizides, J.~Kamin, and D.~d'Enterria, ``Improved {Monte
  Carlo Glauber} predictions at present and future nuclear colliders'',}
  \textit{ Phys. Rev. C} \textbf{ 97} (2018) 054910,
  \href{http://dx.doi.org/10.1103/PhysRevC.97.054910}{\doi{10.1103/PhysRevC.97.054910}},
  \href{http://www.arXiv.org/abs/1710.07098}{\texttt{arXiv:1710.07098}}.
  [Erratum: \DOI{10.1103/PhysRevC.99.019901}].

\bibitem{CMS:2024krd}
\hrefCMSnoop {}{{CMS Collaboration}, ``{Overview of high-density QCD studies
  with the CMS experiment at the LHC}'',} 2024.
  \href{http://www.arXiv.org/abs/2405.10785}{\texttt{arXiv:2405.10785}}.
  Submitted to \textit{Phys. Rep.}

\bibitem{LHCb:2020jse}
\hrefCMSnoop {}{{LHCb Collaboration}, ``Observation of enhanced double parton
  scattering in proton-lead collisions at $\sqrt{s_{_{\mathrm{NN}}}}
  =8.16$~{TeV}'',} \textit{ Phys. Rev. Lett.} \textbf{ 125} (2020) 212001,
  \href{http://dx.doi.org/10.1103/PhysRevLett.125.212001}{\doi{10.1103/PhysRevLett.125.212001}},
  \href{http://www.arXiv.org/abs/2007.06945}{\texttt{arXiv:2007.06945}}.

\bibitem{Lansberg:2019adr}
\hrefCMSnoop {}{J.-P. Lansberg, ``New observables in inclusive production of
  quarkonia'',} \textit{ Phys. Rept.} \textbf{ 889} (2020) 1,
  \href{http://dx.doi.org/10.1016/j.physrep.2020.08.007}{\doi{10.1016/j.physrep.2020.08.007}},
  \href{http://www.arXiv.org/abs/1903.09185}{\texttt{arXiv:1903.09185}}.

\bibitem{Bodwin:1994jh}
\hrefCMSnoop {}{G.~T. Bodwin, E.~Braaten, and G.~P. Lepage, ``Rigorous {QCD}
  analysis of inclusive annihilation and production of heavy quarkonium'',}
  \textit{ Phys. Rev. D} \textbf{ 51} (1995) 1125,
  \href{http://dx.doi.org/10.1103/PhysRevD.55.5853}{\doi{10.1103/PhysRevD.55.5853}},
  \href{http://www.arXiv.org/abs/hep-ph/9407339}{\texttt{arXiv:hep-ph/9407339}}.
  [Erratum: \DOI{10.1103/PhysRevD.55.5853}].

\bibitem{HEPData}
\hrefCMSnoop {}{}{HEPD}ata record for this analysis, 2024.
\newblock
  \href{http://dx.doi.org/10.17182/hepdata.152618}{\doi{10.17182/hepdata.152618}}.

\bibitem{CMS:2008xjf}
\hrefCMSnoop {}{{CMS Collaboration}, ``The {CMS} experiment at the {CERN
  LHC}'',} \textit{ JINST} \textbf{ 3} (2008) S08004,
  \href{http://dx.doi.org/10.1088/1748-0221/3/08/S08004}{\doi{10.1088/1748-0221/3/08/S08004}}.

\bibitem{CMS:2020cmk}
\hrefCMSnoop {}{{CMS Collaboration}, ``Performance of the {CMS} level-1 trigger
  in proton-proton collisions at $\sqrt{s} =13$~{TeV}'',} \textit{ JINST}
  \textbf{ 15} (2020) P10017,
  \href{http://dx.doi.org/10.1088/1748-0221/15/10/P10017}{\doi{10.1088/1748-0221/15/10/P10017}},
  \href{http://www.arXiv.org/abs/2006.10165}{\texttt{arXiv:2006.10165}}.

\bibitem{CMS:2016ngn}
\hrefCMSnoop {}{{CMS Collaboration}, ``The {CMS} trigger system'',} \textit{
  JINST} \textbf{ 12} (2017) P01020,
  \href{http://dx.doi.org/10.1088/1748-0221/12/01/P01020}{\doi{10.1088/1748-0221/12/01/P01020}},
  \href{http://www.arXiv.org/abs/1609.02366}{\texttt{arXiv:1609.02366}}.

\bibitem{CMS:2020uim}
\hrefCMSnoop {}{{CMS Collaboration}, ``Electron and photon reconstruction and
  identification with the {CMS} experiment at the {CERN LHC}'',} \textit{
  JINST} \textbf{ 16} (2021) P05014,
  \href{http://dx.doi.org/10.1088/1748-0221/16/05/P05014}{\doi{10.1088/1748-0221/16/05/P05014}},
  \href{http://www.arXiv.org/abs/2012.06888}{\texttt{arXiv:2012.06888}}.

\bibitem{Sirunyan:2018fpa}
\hrefCMSnoop {}{{CMS Collaboration}, ``Performance of the {CMS} muon detector
  and muon reconstruction with proton-proton collisions at $\sqrt{s}=
  13$~{TeV}'',} \textit{ JINST} \textbf{ 13} (2018) P06015,
  \href{http://dx.doi.org/10.1088/1748-0221/13/06/P06015}{\doi{10.1088/1748-0221/13/06/P06015}},
  \href{http://www.arXiv.org/abs/1804.04528}{\texttt{arXiv:1804.04528}}.

\bibitem{CMS:2014pgm}
\hrefCMSnoop {}{{CMS Collaboration}, ``Description and performance of track and
  primary-vertex reconstruction with the {CMS} tracker'',} \textit{ JINST}
  \textbf{ 9} (2014) P10009,
  \href{http://dx.doi.org/10.1088/1748-0221/9/10/P10009}{\doi{10.1088/1748-0221/9/10/P10009}},
  \href{http://www.arXiv.org/abs/1405.6569}{\texttt{arXiv:1405.6569}}.

\bibitem{CMS:2017yfk}
\hrefCMSnoop {}{{CMS Collaboration}, ``Particle-flow reconstruction and global
  event description with the {CMS} detector'',} \textit{ JINST} \textbf{ 12}
  (2017) P10003,
  \href{http://dx.doi.org/10.1088/1748-0221/12/10/P10003}{\doi{10.1088/1748-0221/12/10/P10003}},
  \href{http://www.arXiv.org/abs/1706.04965}{\texttt{arXiv:1706.04965}}.

\bibitem{CMS:2024qgs}
\hrefCMSnoop {}{{CMS Collaboration}, ``{Performance of CMS muon reconstruction
  from proton-proton to heavy ion collisions}'',} \textit{ JINST} \textbf{ 19}
  (2024) P09012,
  \href{http://dx.doi.org/10.1088/1748-0221/19/09/P09012}{\doi{10.1088/1748-0221/19/09/P09012}},
  \href{http://www.arXiv.org/abs/2404.17377}{\texttt{arXiv:2404.17377}}.

\bibitem{Khachatryan:2016odn}
\hrefCMSnoop {}{{CMS Collaboration}, ``Charged-particle nuclear modification
  factors in {PbPb} and {pPb} collisions at
  $\sqrt{s_{_{\mathrm{NN}}}}=5.02$~{TeV}'',} \textit{ JHEP} \textbf{ 04} (2017)
  039,
  \href{http://dx.doi.org/10.1007/JHEP04(2017)039}{\doi{10.1007/JHEP04(2017)039}},
  \href{http://www.arXiv.org/abs/1611.01664}{\texttt{arXiv:1611.01664}}.

\bibitem{CMS-TDR-15-02}
\href {http://cds.cern.ch/record/2020886}{{CMS Collaboration}, ``Technical
  proposal for the {Phase-II} upgrade of the {Compact Muon Solenoid}'',} CMS
  Technical Proposal CERN-LHCC-2015-010, CMS-TDR-15-02, 2015.

\bibitem{Fruhwirth:1987fm}
\hrefCMSnoop {}{R.~Fr{\"u}hwirth, ``Application of {Kalman} filtering to track
  and vertex fitting'',} \textit{ Nucl. Instrum. Meth. A} \textbf{ 262} (1987)
  444,
  \href{http://dx.doi.org/10.1016/0168-9002(87)90887-4}{\doi{10.1016/0168-9002(87)90887-4}}.

\bibitem{CMS:2017dju}
\hrefCMSnoop {}{{CMS Collaboration}, ``Measurement of quarkonium production
  cross sections in pp collisions at $\sqrt{s}=13$~{TeV}'',} \textit{ Phys.
  Lett. B} \textbf{ 780} (2018) 251,
  \href{http://dx.doi.org/10.1016/j.physletb.2018.02.033}{\doi{10.1016/j.physletb.2018.02.033}},
  \href{http://www.arXiv.org/abs/1710.11002}{\texttt{arXiv:1710.11002}}.

\bibitem{ATLAS:2011aqv}
\hrefCMSnoop {}{{ATLAS Collaboration}, ``{Measurement of the differential
  cross-sections of inclusive, prompt and non-prompt $J/\psi$ production in
  proton-proton collisions at $\sqrt{s}=7$~TeV}'',} \textit{ Nucl. Phys. B}
  \textbf{ 850} (2011) 387,
  \href{http://dx.doi.org/10.1016/j.nuclphysb.2011.05.015}{\doi{10.1016/j.nuclphysb.2011.05.015}},
  \href{http://www.arXiv.org/abs/1104.3038}{\texttt{arXiv:1104.3038}}.

\bibitem{Sjostrand:2014zea}
T.~Sj{\"o}strand\hrefCMSnoop {}{ { et~al.}, ``An introduction to {PYTHIA
  8.2}'',} \textit{ Comput. Phys. Commun.} \textbf{ 191} (2015) 159,
  \href{http://dx.doi.org/10.1016/j.cpc.2015.01.024}{\doi{10.1016/j.cpc.2015.01.024}},
  \href{http://www.arXiv.org/abs/1410.3012}{\texttt{arXiv:1410.3012}}.

\bibitem{Pierog:2013ria}
T.~Pierog\hrefCMSnoop {}{ { et~al.}, ``{EPOS LHC}: Test of collective
  hadronization with data measured at the {CERN Large Hadron Collider}'',}
  \textit{ Phys. Rev. C} \textbf{ 92} (2015) 034906,
  \href{http://dx.doi.org/10.1103/PhysRevC.92.034906}{\doi{10.1103/PhysRevC.92.034906}},
  \href{http://www.arXiv.org/abs/1306.0121}{\texttt{arXiv:1306.0121}}.

\bibitem{CMS:2013gbz}
\hrefCMSnoop {}{{CMS Collaboration}, ``Measurement of the prompt {\JPsi} and
  $\psi${(2S)} polarizations in pp collisions at $\sqrt{s} = 7$~{TeV}'',}
  \textit{ Phys. Lett. B} \textbf{ 727} (2013) 381,
  \href{http://dx.doi.org/10.1016/j.physletb.2013.10.055}{\doi{10.1016/j.physletb.2013.10.055}},
  \href{http://www.arXiv.org/abs/1307.6070}{\texttt{arXiv:1307.6070}}.

\bibitem{LHCb:2013izl}
\hrefCMSnoop {}{{LHCb Collaboration}, ``Measurement of {\JPsi} polarization in
  pp collisions at $\sqrt{s}=7$~{TeV}'',} \textit{ Eur. Phys. J. C} \textbf{
  73} (2013) 2631,
  \href{http://dx.doi.org/10.1140/epjc/s10052-013-2631-3}{\doi{10.1140/epjc/s10052-013-2631-3}},
  \href{http://www.arXiv.org/abs/1307.6379}{\texttt{arXiv:1307.6379}}.

\bibitem{ALICE:2020iev}
\hrefCMSnoop {}{{ALICE Collaboration}, ``First measurement of quarkonium
  polarization in nuclear collisions at the {LHC}'',} \textit{ Phys. Lett. B}
  \textbf{ 815} (2021) 136146,
  \href{http://dx.doi.org/10.1016/j.physletb.2021.136146}{\doi{10.1016/j.physletb.2021.136146}},
  \href{http://www.arXiv.org/abs/2005.11128}{\texttt{arXiv:2005.11128}}.

\bibitem{Agostinelli:2002hh}
\hrefCMSnoop {}{{GEANT4} Collaboration, ``{GEANT4} --- a simulation toolkit'',}
  \textit{ Nucl. Instrum. Meth. A} \textbf{ 506} (2003) 250,
  \href{http://dx.doi.org/10.1016/S0168-9002(03)01368-8}{\doi{10.1016/S0168-9002(03)01368-8}}.

\bibitem{Oreglia:1980cs}
\href {http://www.slac.stanford.edu/pubs/slacreports/slac-r-236.html}{M.~J.
  Oreglia, ``{A study of the reactions $\psi^\prime \to \gamma \gamma
  \psi$}''}.
\newblock PhD thesis, Stanford University, 1980.
\newblock {SLAC} Report {SLAC-R-236}.

\bibitem{ParticleDataGroup:2022pth}
\hrefCMSnoop {}{{Particle Data Group} Collaboration, ``Review of particle
  physics'',} \textit{ PTEP} \textbf{ 2022} (2022) 083C01,
  \href{http://dx.doi.org/10.1093/ptep/ptac097}{\doi{10.1093/ptep/ptac097}}.

\bibitem{Cowan:2010js}
\hrefCMSnoop {}{G.~Cowan, K.~Cranmer, E.~Gross, and O.~Vitells, ``Asymptotic
  formulae for likelihood-based tests of new physics'',} \textit{ Eur. Phys. J.
  C} \textbf{ 71} (2011) 1554,
  \href{http://dx.doi.org/10.1140/epjc/s10052-011-1554-0}{\doi{10.1140/epjc/s10052-011-1554-0}},
  \href{http://www.arXiv.org/abs/1007.1727}{\texttt{arXiv:1007.1727}}.
  [Erratum: \DOI{10.1140/epjc/s10052-013-2501-z}].

\bibitem{Wilks:1938dza}
\hrefCMSnoop {}{S.~S. Wilks, ``The large-sample distribution of the likelihood
  ratio for testing composite hypotheses'',} \textit{ Annals Math. Statist.}
  \textbf{ 9} (1938) 60,
  \href{http://dx.doi.org/10.1214/aoms/1177732360}{\doi{10.1214/aoms/1177732360}}.

\bibitem{CMS-DP-2020-021}
\href {https://cds.cern.ch/record/2717925}{{CMS Collaboration}, ``{ECAL} 2016
  refined calibration and {Run2} summary plots'',} CMS Detector Performance
  Note CMS-DP-2020-021, 2020.

\bibitem{679b84c4-7042-35e5-aa82-ed196e957f94}
\hrefCMSnoop {}{F.~Mosteller and R.~A. Fisher, ``Questions and answers'',}
  \textit{ Amer. Statis.} \textbf{ 2} (1948) 30,
  \href{http://dx.doi.org/10.2307/2681650}{\doi{10.2307/2681650}}.

\bibitem{Pivk:2004ty}
\hrefCMSnoop {}{M.~Pivk and F.~R. Le~Diberder, ``{$_{s}Plot$}: {A} statistical
  tool to unfold data distributions'',} \textit{ Nucl. Instrum. Meth. A}
  \textbf{ 555} (2005) 356,
  \href{http://dx.doi.org/10.1016/j.nima.2005.08.106}{\doi{10.1016/j.nima.2005.08.106}},
  \href{http://www.arXiv.org/abs/physics/0402083}{\texttt{arXiv:physics/0402083}}.

\bibitem{CMS:2021xjt}
\hrefCMSnoop {}{{CMS Collaboration}, ``Precision luminosity measurement in
  proton-proton collisions at $\sqrt{s} =$ 13~{TeV} in 2015 and 2016 at
  {CMS}'',} \textit{ Eur. Phys. J. C} \textbf{ 81} (2021) 800,
  \href{http://dx.doi.org/10.1140/epjc/s10052-021-09538-2}{\doi{10.1140/epjc/s10052-021-09538-2}},
  \href{http://www.arXiv.org/abs/2104.01927}{\texttt{arXiv:2104.01927}}.

\bibitem{CMS-PAS-LUM-18-002}
\href {https://cds.cern.ch/record/2676164/}{{CMS Collaboration}, ``{CMS}
  luminosity measurement for the 2018 data-taking period at $\sqrt{s} =
  13$~{TeV}'',} CMS Physics Analysis Summary CMS-PAS-LUM-18-002, 2019.

\bibitem{Shao:2012iz}
\hrefCMSnoop {}{H.-S. Shao, ``{HELAC-Onia}: An automatic matrix element
  generator for heavy quarkonium physics'',} \textit{ Comput. Phys. Commun.}
  \textbf{ 184} (2013) 2562,
  \href{http://dx.doi.org/10.1016/j.cpc.2013.05.023}{\doi{10.1016/j.cpc.2013.05.023}},
  \href{http://www.arXiv.org/abs/1212.5293}{\texttt{arXiv:1212.5293}}.

\bibitem{Shao:2015vga}
\hrefCMSnoop {}{H.-S. Shao, ``{HELAC-Onia 2.0}: an upgraded matrix-element and
  event generator for heavy quarkonium physics'',} \textit{ Comput. Phys.
  Commun.} \textbf{ 198} (2016) 238,
  \href{http://dx.doi.org/10.1016/j.cpc.2015.09.011}{\doi{10.1016/j.cpc.2015.09.011}},
  \href{http://www.arXiv.org/abs/1507.03435}{\texttt{arXiv:1507.03435}}.

\bibitem{Dulat:2015mca}
S.~Dulat\hrefCMSnoop {}{ { et~al.}, ``New parton distribution functions from a
  global analysis of quantum chromodynamics'',} \textit{ Phys. Rev. D} \textbf{
  93} (2016) 033006,
  \href{http://dx.doi.org/10.1103/PhysRevD.93.033006}{\doi{10.1103/PhysRevD.93.033006}},
  \href{http://www.arXiv.org/abs/1506.07443}{\texttt{arXiv:1506.07443}}.

\bibitem{Eskola:2016oht}
\hrefCMSnoop {}{K.~J. Eskola, P.~Paakkinen, H.~Paukkunen, and C.~A. Salgado,
  ``{EPPS16}: Nuclear parton distributions with {LHC} data'',} \textit{ Eur.
  Phys. J. C} \textbf{ 77} (2017) 163,
  \href{http://dx.doi.org/10.1140/epjc/s10052-017-4725-9}{\doi{10.1140/epjc/s10052-017-4725-9}},
  \href{http://www.arXiv.org/abs/1612.05741}{\texttt{arXiv:1612.05741}}.

\bibitem{Kusina:2017gkz}
\hrefCMSnoop {}{A.~Kusina, J.-P. Lansberg, I.~Schienbein, and H.-S. Shao,
  ``Gluon shadowing in heavy-flavor production at the {LHC}'',} \textit{ Phys.
  Rev. Lett.} \textbf{ 121} (2018) 052004,
  \href{http://dx.doi.org/10.1103/PhysRevLett.121.052004}{\doi{10.1103/PhysRevLett.121.052004}},
  \href{http://www.arXiv.org/abs/1712.07024}{\texttt{arXiv:1712.07024}}.

\bibitem{Kusina:2020dki}
\hrefCMSnoop {}{A.~Kusina, J.-P. Lansberg, I.~Schienbein, and H.-S. Shao,
  ``Reweighted nuclear {PDFs} using heavy-flavor production data at the
  {LHC}'',} \textit{ Phys. Rev. D} \textbf{ 104} (2021) 014010,
  \href{http://dx.doi.org/10.1103/PhysRevD.104.014010}{\doi{10.1103/PhysRevD.104.014010}},
  \href{http://www.arXiv.org/abs/2012.11462}{\texttt{arXiv:2012.11462}}.

\bibitem{Lansberg:2016deg}
\hrefCMSnoop {}{J.-P. Lansberg and H.-S. Shao, ``Towards an automated tool to
  evaluate the impact of the nuclear modification of the gluon density on
  quarkonium, {D} and {B} meson production in proton\textendash{}nucleus
  collisions'',} \textit{ Eur. Phys. J. C} \textbf{ 77} (2017) 1,
  \href{http://dx.doi.org/10.1140/epjc/s10052-016-4575-x}{\doi{10.1140/epjc/s10052-016-4575-x}},
  \href{http://www.arXiv.org/abs/1610.05382}{\texttt{arXiv:1610.05382}}.

\bibitem{Lansberg:2013qka}
\hrefCMSnoop {}{J.-P. Lansberg and H.-S. Shao, ``Production of {J}$/\psi +
  \eta_{c}$ versus {J}$/\psi + \mathrm{J}/\psi$ at the {LHC}: Importance of
  real $\alpha^{5}_\mathrm{s}$ corrections'',} \textit{ Phys. Rev. Lett.}
  \textbf{ 111} (2013) 122001,
  \href{http://dx.doi.org/10.1103/PhysRevLett.111.122001}{\doi{10.1103/PhysRevLett.111.122001}},
  \href{http://www.arXiv.org/abs/1308.0474}{\texttt{arXiv:1308.0474}}.

\bibitem{Lansberg:2014swa}
\hrefCMSnoop {}{J.-P. Lansberg and H.-S. Shao, ``{J}/\ensuremath{\psi} -pair
  production at large momenta: {Indications} for double parton scatterings and
  large \ensuremath{\alpha}$_s^5$ contributions'',} \textit{ Phys. Lett. B}
  \textbf{ 751} (2015) 479,
  \href{http://dx.doi.org/10.1016/j.physletb.2015.10.083}{\doi{10.1016/j.physletb.2015.10.083}},
  \href{http://www.arXiv.org/abs/1410.8822}{\texttt{arXiv:1410.8822}}.

\bibitem{Lansberg:2017chq}
\hrefCMSnoop {}{J.-P. Lansberg, H.-S. Shao, and N.~Yamanaka, ``Indication for
  double parton scatterings in {W}$^+$ prompt {J}/$\psi$ production at the
  {LHC}'',} \textit{ Phys. Lett. B} \textbf{ 781} (2018) 485,
  \href{http://dx.doi.org/10.1016/j.physletb.2018.04.020}{\doi{10.1016/j.physletb.2018.04.020}},
  \href{http://www.arXiv.org/abs/1707.04350}{\texttt{arXiv:1707.04350}}.

\bibitem{ATLAS:2014ofp}
\hrefCMSnoop {}{{ATLAS Collaboration}, ``Observation and measurements of the
  production of prompt and non-prompt {J}$/\psi$ mesons in association with a
  {Z} boson in pp collisions at $\sqrt{s}= 8$~{TeV} with the {ATLAS}
  detector'',} \textit{ Eur. Phys. J. C} \textbf{ 75} (2015) 229,
  \href{http://dx.doi.org/10.1140/epjc/s10052-015-3406-9}{\doi{10.1140/epjc/s10052-015-3406-9}},
  \href{http://www.arXiv.org/abs/1412.6428}{\texttt{arXiv:1412.6428}}.

\bibitem{Shao:2016wor}
\hrefCMSnoop {}{H.-S. Shao and Y.-J. Zhang, ``Complete study of hadroproduction
  of a {$\Upsilon$} meson associated with a prompt {J}$/\psi$'',} \textit{
  Phys. Rev. Lett.} \textbf{ 117} (2016) 062001,
  \href{http://dx.doi.org/10.1103/PhysRevLett.117.062001}{\doi{10.1103/PhysRevLett.117.062001}},
  \href{http://www.arXiv.org/abs/1605.03061}{\texttt{arXiv:1605.03061}}.

\bibitem{Lansberg:2016muq}
\hrefCMSnoop {}{J.-P. Lansberg and H.-S. Shao, ``Phenomenological analysis of
  associated production of {$\PZ^0+\mathrm{b}$ in the $\mathrm{b} \to \JPsi
  \mathrm{X}$} decay channel at the {LHC}'',} \textit{ Nucl. Phys. B} \textbf{
  916} (2017) 132,
  \href{http://dx.doi.org/10.1016/j.nuclphysb.2017.01.002}{\doi{10.1016/j.nuclphysb.2017.01.002}},
  \href{http://www.arXiv.org/abs/1611.09303}{\texttt{arXiv:1611.09303}}.

\bibitem{UA2:1991apc}
\hrefCMSnoop {}{{UA2} Collaboration, ``A study of multi-jet events at the
  {CERN} $\mathrm{\overline{p}p}$ collider and a search for double parton
  scattering'',} \textit{ Phys. Lett. B} \textbf{ 268} (1991) 145,
  \href{http://dx.doi.org/10.1016/0370-2693(91)90937-L}{\doi{10.1016/0370-2693(91)90937-L}}.

\bibitem{CDF:1993sbj}
\hrefCMSnoop {}{{CDF} Collaboration, ``Study of four jet events and evidence
  for double parton interactions in $\mathrm{\overline{p}p}$ collisions at
  $\sqrt{s} = 1.8$~{TeV}'',} \textit{ Phys. Rev. D} \textbf{ 47} (1993) 4857,
  \href{http://dx.doi.org/10.1103/PhysRevD.47.4857}{\doi{10.1103/PhysRevD.47.4857}}.

\bibitem{CDF:1997yfa}
\hrefCMSnoop {}{{CDF} Collaboration, ``Double parton scattering in
  $\mathrm{\overline{p}p}$ collisions at $\sqrt{s} = 1.8$~{TeV}'',} \textit{
  Phys. Rev. D} \textbf{ 56} (1997) 3811,
  \href{http://dx.doi.org/10.1103/PhysRevD.56.3811}{\doi{10.1103/PhysRevD.56.3811}}.

\bibitem{D0:2009apj}
\hrefCMSnoop {}{{D0} Collaboration, ``Double parton interactions in $\gamma$+3
  jet events in $\mathrm{\overline{p}p}$ collisions $\sqrt{s}=1.96$~{TeV}'',}
  \textit{ Phys. Rev. D} \textbf{ 81} (2010) 052012,
  \href{http://dx.doi.org/10.1103/PhysRevD.81.052012}{\doi{10.1103/PhysRevD.81.052012}},
  \href{http://www.arXiv.org/abs/0912.5104}{\texttt{arXiv:0912.5104}}.

\bibitem{Khachatryan:2015pea}
\hrefCMSnoop {}{{CMS Collaboration}, ``Event generator tunes obtained from
  underlying event and multiparton scattering measurements'',} \textit{ Eur.
  Phys. J. C} \textbf{ 76} (2016) 155,
  \href{http://dx.doi.org/10.1140/epjc/s10052-016-3988-x}{\doi{10.1140/epjc/s10052-016-3988-x}},
  \href{http://www.arXiv.org/abs/1512.00815}{\texttt{arXiv:1512.00815}}.

\bibitem{D0:2015rpo}
\hrefCMSnoop {}{{D0} Collaboration, ``{Study of double parton interactions in
  diphoton + dijet events in $\mathrm{\overline{p}p}$ collisions at $\sqrt{s} =
  1.96$~TeV}'',} \textit{ Phys. Rev. D} \textbf{ 93} (2016) 052008,
  \href{http://dx.doi.org/10.1103/PhysRevD.93.052008}{\doi{10.1103/PhysRevD.93.052008}},
  \href{http://www.arXiv.org/abs/1512.05291}{\texttt{arXiv:1512.05291}}.

\bibitem{D0:2014owy}
\hrefCMSnoop {}{{D0} Collaboration, ``{Double Parton Interactions in $\gamma+3$
  jet and $\gamma+\mathrm{b/c}$ jet + 2 jet events in $\mathrm{\overline{p}p}$
  collisions at $\sqrt{s} = 1.96$~TeV}'',} \textit{ Phys. Rev. D} \textbf{ 89}
  (2014) 072006,
  \href{http://dx.doi.org/10.1103/PhysRevD.89.072006}{\doi{10.1103/PhysRevD.89.072006}},
  \href{http://www.arXiv.org/abs/1402.1550}{\texttt{arXiv:1402.1550}}.

\end{thebibliography}\endgroup
\cleardoublepage \appendix\section{The CMS Collaboration \label{app:collab}}\begin{sloppypar}\hyphenpenalty=5000\widowpenalty=500\clubpenalty=5000
\cmsinstitute{Yerevan Physics Institute, Yerevan, Armenia}
{\tolerance=6000
A.~Hayrapetyan, A.~Tumasyan\cmsAuthorMark{1}\cmsorcid{0009-0000-0684-6742}
\par}
\cmsinstitute{Institut f\"{u}r Hochenergiephysik, Vienna, Austria}
{\tolerance=6000
W.~Adam\cmsorcid{0000-0001-9099-4341}, J.W.~Andrejkovic, T.~Bergauer\cmsorcid{0000-0002-5786-0293}, S.~Chatterjee\cmsorcid{0000-0003-2660-0349}, K.~Damanakis\cmsorcid{0000-0001-5389-2872}, M.~Dragicevic\cmsorcid{0000-0003-1967-6783}, P.S.~Hussain\cmsorcid{0000-0002-4825-5278}, M.~Jeitler\cmsAuthorMark{2}\cmsorcid{0000-0002-5141-9560}, N.~Krammer\cmsorcid{0000-0002-0548-0985}, A.~Li\cmsorcid{0000-0002-4547-116X}, D.~Liko\cmsorcid{0000-0002-3380-473X}, I.~Mikulec\cmsorcid{0000-0003-0385-2746}, J.~Schieck\cmsAuthorMark{2}\cmsorcid{0000-0002-1058-8093}, R.~Sch\"{o}fbeck\cmsorcid{0000-0002-2332-8784}, D.~Schwarz\cmsorcid{0000-0002-3821-7331}, M.~Sonawane\cmsorcid{0000-0003-0510-7010}, S.~Templ\cmsorcid{0000-0003-3137-5692}, W.~Waltenberger\cmsorcid{0000-0002-6215-7228}, C.-E.~Wulz\cmsAuthorMark{2}\cmsorcid{0000-0001-9226-5812}
% ...
\par}
\vskip\cmsinstskip
\dag:~Deceased\\
$^{1}$Also at Yerevan State University, Yerevan, Armenia\\
$^{2}$Also at TU Wien, Vienna, Austria\\
% ...
\end{sloppypar}
\end{document}